\documentclass[journal,twoside]{IEEEtran}
\usepackage{amssymb}
\usepackage{amsthm}
\usepackage{multirow}
\usepackage{inputenc}
\usepackage{enumerate} 
\usepackage{textcomp}
\usepackage{listings}
\usepackage{array}
\usepackage[switch,mathlines]{lineno}
\usepackage{soul}
\usepackage{xfrac}
\usepackage{graphicx}
\usepackage{cite}
\usepackage[cmex10]{amsmath}
\usepackage{algorithm}
\usepackage{algorithmic}
\usepackage{cancel}
\usepackage{makecell}
\usepackage{pifont}
\usepackage{tikz}
\usepackage{pgfplots}
\pgfplotsset{compat=1.18}
\usetikzlibrary{calc,arrows.meta}
\definecolor{initfill}{RGB}{255,246,218}
\definecolor{varfill}{RGB}{219,235,247}
\definecolor{measpurple}{RGB}{139,77,173}
\usepackage{stackengine}

\usepackage{CJK}
\usepackage[cmex10]{amsmath}
\usepackage{bm}
\usepackage{color}
\usepackage{xcolor}
\usepackage{amsmath,amsthm,amssymb,amsfonts}
\usepackage{flushend}
\usepackage{float}

\usepackage{arydshln} 
\usepackage{hyperref}
\hypersetup{
     colorlinks   = true,
     linkcolor    = blue,
     citecolor    = red,
     urlcolor     = blue
}
\makeatletter
\newcommand*{\transpose}{%
  {\mathpalette\@transpose{}}%
}
\newcommand*{\@transpose}[2]{%
  \raisebox{\depth}{$\m@th#1\intercal$}%
}
\makeatother

\usepackage{soul}
\usepackage{tikz}
\usepackage{booktabs}
\usepackage{tabularx}

\usepackage[caption=false,font=footnotesize]{subfig}
\ifCLASSINFOpdf
\else
\fi

\begin{document}
\renewcommand{\ttdefault}{cmtt}
\bstctlcite{IEEEexample:BSTcontrol}

\title{Quantum Computing in Power Systems: \\ Current Landscape and Future Roadmap with AI}



\author{
{Yijun~Xu,~\IEEEmembership{Senior Member, IEEE},
Yongbing Yao,~\IEEEmembership{Student Member, IEEE},
Wei Gu,~\IEEEmembership{Senior Member, IEEE},
Chengjun Liu,
Qi Lou,
Lamine Mili,~\IEEEmembership{Life Fellow, IEEE},
Mohammadreza Saghafi, 
Mert Korkali,~\IEEEmembership{Senior Member, IEEE}, and
Shuai Lu
}
\thanks{Yijun Xu, Yongbing Yao, Wei Gu, Chengjun Liu, Qi Lou, and Shuai Lu are with the School of Electrical Engineering, Southeast University, Nanjing 210096, China (e-mail: yijunxu@seu.edu.cn; yongbingyao@seu.edu.cn; wgu@seu.edu.cn; liucj@seu.edu.cn; louqi@seu.edu.cn; shuai.lu.seu@outlook.com).}
\thanks{Lamine Mili and Mohammadreza Saghafi are with the Bradley Department of Electrical and Computer Engineering, Virginia Tech, Northern Virginia Center, Falls Church, VA 22043 USA (e-mail: lmili@vt.edu; saghafim@vt.edu).}
\thanks{Mert Korkali is with the Department of Electrical Engineering and Computer Science, University of Missouri, Columbia, MO 65211 USA (e-mail: korkalim@missouri.edu).}
\thanks{(Corresponding author: Wei Gu)}
}

\markboth{}%
{Xu \MakeLowercase{\textit{\textit{et al.}}}: Quantum Computing in Power Systems: Current Landscape and Future Roadmap with AI}
\maketitle
\begin{abstract}
Modern power systems are characterized by greater penetration of renewable energy resources, extensive use of power electronics, and more diverse loads. 
This leads to a dramatic increase in computational complexity, which, in turn, compromises solution efficiency and fails to satisfy the real-time demands of system analysis. 
Recently, quantum computing techniques, attributed to their unique physical properties, such as superposition and entanglement, hold the potential to offer new solutions for power systems. 
To this end, this paper not only provides a comprehensive landscape of quantum computing in power systems, but also investigates its roadmap with AI-enabled evolution. 
This paper begins with an introduction to the fundamental concepts of quantum computing, followed by a review of several approaches to power system simulation and optimization. 
Subsequently, we further investigate the potential quantum AI-enabled evolution in power systems. 
In addition, the current development of quantum software and hardware platforms is elaborated. 
Finally, this paper synthesizes the prevailing challenges and prospects for future research directions.
\end{abstract}

\IEEEpeerreviewmaketitle

\vspace{-0.4cm}
\section{Introduction}
\IEEEPARstart{W}{ith} the rapid development of modern power systems, it exhibits significant characteristics, such as increasing penetration of renewable generation, diversified load demand, and widespread power electronics devices \cite{10799930}. 
These trends drastically increase computational complexity, leading to heavy computational burdens that challenge classical computers constrained by Moore’s Law \cite{9275593}.  

Currently, modern power systems impose real-time requirements on certain computational tasks \cite{10419164}, such as electromagnetic simulation and optimization. 
For accurate simulation of power electronic converters, simulation step sizes must be set in the range of hundreds of nanoseconds to a few microseconds \cite{beg2018multi}. 
However, solving the linear equations corresponding to high-dimensional admittance matrices can be extremely time-consuming, limited by the computational capabilities of classical computers. 
This severely constrains the realization of large-scale electromagnetic transient simulations in real time. 
For optimization, the dispatch model of a power system contains tens of thousands of discrete decision variables and intricate constraints \cite{barros2025globally}, making it difficult to obtain an optimal solution in real time. 
This can result in substantial economic losses. 
Overall, the contradiction between computational complexity and real-time operation is becoming increasingly pronounced. 
Moreover, given the highly intricate structures and nonlinear features of power system data, classical AI algorithms, which are limited in expressive power, struggle to handle such data. 
These obstacles will hinder the development of modern power systems. 

In this context, there is a pressing need for efficient computational paradigms. 
In recent years, the rapid advancement of quantum computing technology has opened up a new avenue to address the challenges described above \cite{pfaendler2024advancements}. 
By leveraging quantum-mechanical properties such as superposition, entanglement, and coherence, quantum computing has achieved revolutionary breakthroughs at the algorithmic level. 
Research indicates that the complexity of solving linear equations can be reduced from the polynomial level on classical computers to the algorithmic level by quantum computing \cite{cai2013experimental}. 
Additionally, it can solve certain NP-hard problems in power systems that are intractable for classical computers. 
In addition, quantum states inherently appear in a high-dimensional Hilbert space, whose dimension increases exponentially with the number of qubits. 
This endows quantum AI with exceptional representational capabilities, enabling more efficient processing of power system data. 
In sum, the adoption of quantum computing techniques has the potential to overcome the limitations of classical computing frameworks in complexity and data-processing capabilities, providing a solid technical foundation for the secure, stable and economical operation of modern power systems. 

\begin{figure*}[htbp]
    \centering
    \includegraphics[width=\linewidth]{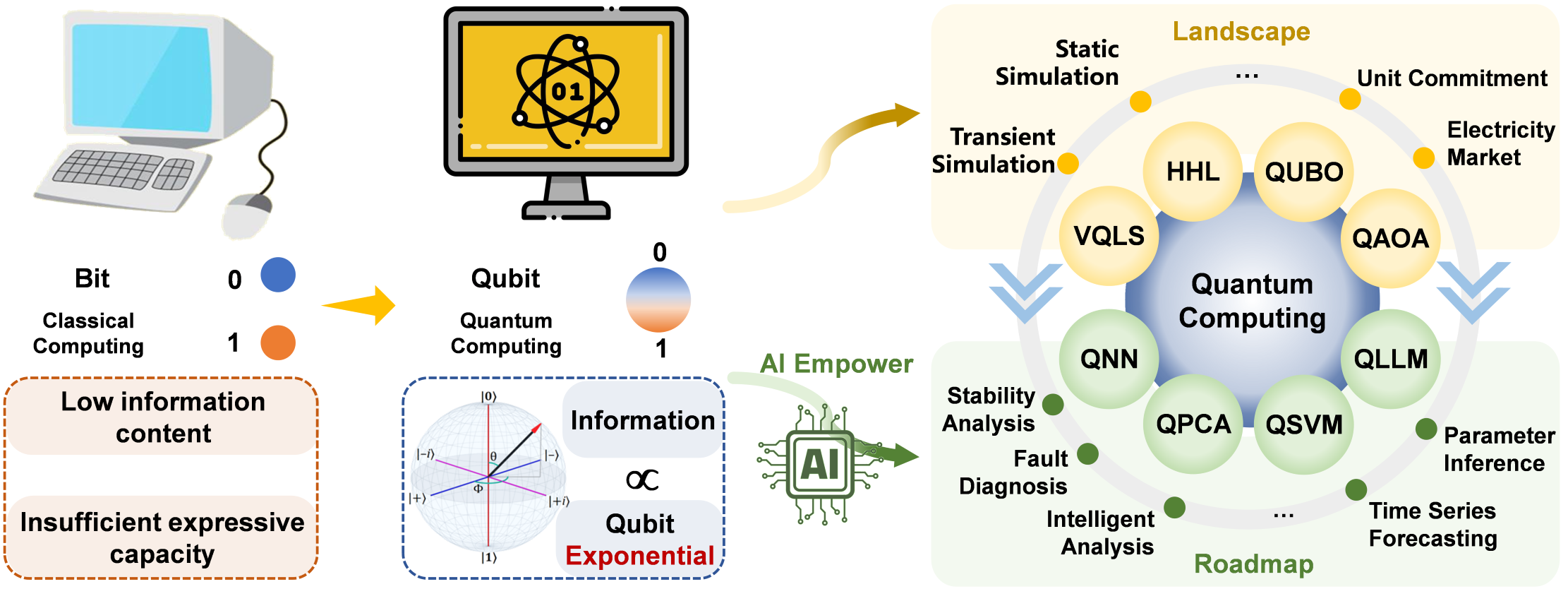}
    \caption{In classical computing, each bit can only take either $0$ or $1$, which leads to a significant growth in computational complexity when dealing with the high-dimensional topologies and massive data in modern power systems. Leveraging unique quantum-mechanical properties, a system composed of $n$ qubits can represent information in a $2^n$-dimensional space, thereby enabling more efficient high-dimensional feature representation and accelerated parallel computation. Leveraging this advantage, researchers have attempted to apply quantum computing techniques to problems such as simulation and optimization in power systems. In recent years, with the rapid advancement of AI technologies, the mutual empowerment of quantum computing and AI has emerged as a research hotspot, with the potential to unlock greater capabilities in the power system domain.}
    \label{fig1}
\end{figure*}

Therefore, this paper focuses on quantum computing technology and provides a systematic discussion of its application in power systems, as shown in Fig. \ref{fig1}. 
Specifically, this includes power system simulations and optimizations, along with the corresponding quantum solution algorithms.

Unlike other quantum review articles, this paper emphasizes a roadmap for integrating AI across multiple dimensions, including quantum deep learning, quantum supervised learning, quantum probabilistic inference, quantum large language models, and more. 
Furthermore, we analyze the potential applications of quantum AI fusion technology in power systems, including fault diagnosis, data mining, and stability assessment, thus establishing meaningful directions for future research. 

The paper is organized as follows: Section II briefly introduces the fundamentals of quantum computing. 
Section III reviews the applications of quantum computing in power system simulation and optimization, followed by Section IV, which is dedicated to the potential quantum AI-enabled evolution in power systems. 
Section V presents a survey of quantum computing hardware and simulation platforms. 
Subsequently, Section VI discusses the key challenges and future research directions. 
In the end, Section VII concludes the paper.

\section{Quantum Computing Techniques}
In this section, we provide a brief introduction to the fundamental concepts of quantum computing and several widely adopted algorithms.

\subsection{Fundamentals of Quantum Computing}
Quantum computing is an emerging computational paradigm built upon the fundamental principles of quantum mechanics. 
Its core idea is to explore the properties of superposition, entanglement, and interference inherent in quantum systems for information processing, in order to overcome the limitations of classical computing in terms of parallelism and computational complexity \cite{nielsen2010quantum}. 

\subsubsection{Mathematical Definition}
In classical computing, the basic unit of information is the bit, which can take only the values $0$ or $1$. 
In contrast, quantum computation generalizes this notion to the qubit, whose state can be expressed as $|\psi\rangle = \alpha |0\rangle + \beta |1\rangle$, where $\alpha$ and $\beta$ are complex probability amplitudes satisfying the normalization condition $|\alpha|^2+|\beta|^2=1$. 
This linear superposition allows a single qubit to represent multiple classical states simultaneously, forming the basis for the intrinsic parallelism of quantum computing. 

\subsubsection{Quantum Superposition and Quantum Entanglement}
When multiple qubits interact, their joint quantum state may no longer be expressible as the product of individual qubit states. 
This gives rise to quantum entanglement, an inseparable composite state such as the Bell state $|\phi\rangle = (|00\rangle + |11\rangle)/\sqrt{2}$, in which the measurement outcome of one qubit instantly determines the corresponding outcome of the other. 
Entanglement reveals the nonlocal correlations of quantum systems and underpins efficient quantum information transfer, quantum teleportation, and quantum error correction. 
Through the combined effects of superposition and entanglement, quantum systems can evolve in an exponentially large Hilbert space, enabling quantum algorithms to outperform their classical counterparts for certain tasks. 

\subsubsection{Quantum Computing Principles}
At present, general quantum computing is predominantly implemented within the quantum gate-based circuit paradigm, in which computation is performed via a sequence of elementary quantum logic gates. 
Mathematically, these gates correspond to unitary transformations that apply reversible operations to quantum states while preserving coherence. 
Representative single-qubit and multi-qubit gates collectively enable state superposition, phase manipulation, and entanglement, thereby providing the fundamental building blocks for expressive quantum algorithms. 
By arranging quantum gates into structured sequences, quantum circuits can be formed to describe the entire computational workflow, including state preparation, quantum evolution, and final measurement \cite{humble2019quantum}. 
The circuit model has thus become the dominant framework for quantum algorithm design and hardware implementation. 

In addition to the gate-based paradigm, several alternative approaches to quantum computing have been explored. 
For example, specialized architectures such as coherent Ising machines and quantum annealers exploit physical energy minimization to solve combinatorial optimization problems. 
In contrast, analog and hybrid quantum simulators provide task-oriented platforms that may offer advantages for specific application domains despite their limited programmability. 

\subsubsection{Quantum Measurement}
Quantum measurement converts a quantum state into observable classical outcomes \cite{braginsky1995quantum}. 
For the quantum state $|\psi\rangle$, the measurement result collapses to $|0\rangle$ with probability $|\alpha|^2$ and to $|1\rangle$ with probability $|\beta|^2$. 
This process is inherently probabilistic and irreversibly collapses the quantum state, thereby terminating its coherent evolution. 
Consequently, quantum algorithms must exploit the advantages of quantum parallelism, such as superposition and interference, before measurement occurs, ensuring that the computation extracts as much useful information as possible prior to this irreversible step. 

Taken together, the theoretical framework of quantum computing achieves exponential state-space expansion through superposition, establishes nonlocal correlations across qubits via entanglement, and enables controlled quantum-state manipulation through quantum gates and circuits. 
Quantum interference and measurement jointly determine the probabilistic nature of computational outcomes. 
These mechanisms allow quantum computing to deliver substantial performance advantages in areas such as factorization, optimization, and quantum simulation \cite{roushan2020quantum}. 
Its inherent parallelism and high-dimensional feature-representation capabilities offer opportunities to overcome computational bottlenecks in power system simulation, optimization, and intelligent data-driven analysis \cite{zhou2022quantum}. 

\subsection{Quantum Linear Solvers}
Solving large-scale linear systems of equations is a core task in scientific computing, motivating the development of quantum algorithms with potential computational advantages over classical methods. 
In this context, the Harrow-Hassidim-Lloyd (HHL) algorithm was proposed as a pioneering quantum approach to solving linear systems under a set of idealized assumptions \cite{harrow2009quantum}, including matrix sparsity, a bounded condition number, efficient quantum state preparation, and the availability of fault-tolerant quantum hardware. 
HHL prepares a quantum state proportional to the solution vector with polylogarithmic complexity in the system dimension. 
Its core mechanism relies on quantum phase estimation to encode eigenvalue information, followed by controlled rotations to invert eigenvalues and reconstruct the solution state \cite{zaman2023step}. 
The detailed derivation of the quantum phase estimation is presented below. 
Given a unitary operator $U$, suppose that it admits an eigenstate $|u\rangle$ satisfying 
\begin{equation}
    U |u\rangle = e ^ {i2\pi \varphi} |u\rangle, 
    \label{qpe}
\end{equation}
where $\varphi \in [0,1)$ is the phase to be estimated. 
Quantum phase estimation aims to encode $\varphi$ into a binary representation stored in a quantum register. 
Assuming that a $t$-qubit register is employed for phase estimation, the initial state of the quantum system is $|\psi_0\rangle = |0\rangle ^ {\otimes t} |u\rangle$. 
First, Hadamard gates are applied to all qubits in the phase register. 
The quantum state becomes 
\begin{equation}
    |\psi_1\rangle = \frac{1}{\sqrt{2^t}} \sum_{k=0} ^ {2^t-1} |k\rangle|u\rangle.
\end{equation}
Subsequently, the qubits in the phase register are employed as control qubits to implement a sequence of controlled unitary operations, $U^{2^0},U^{2^1},...,U^{2^{t-1}}$, in the eigenstate register. 
According to \eqref{qpe}, the quantum state is 
\begin{equation}
\begin{aligned}
    |\psi_2\rangle = &\frac{1}{\sqrt{2^t}}(|0\rangle + e^{i2\pi 2^{t-1}\varphi}|1\rangle)(|0\rangle + e^{i2\pi 2^{t-2}\varphi}|1\rangle)...\\&(|0\rangle + e^{i2\pi 2^0\varphi}|1\rangle)|u\rangle \\
    = & \frac{1}{\sqrt{2^t}} \sum _{k=0}^{2^t-1} e^{i2\pi \varphi k}|k\rangle |u\rangle.
    \label{qpe1}
\end{aligned}
\end{equation}
Assume that the phase, $\varphi$, can be expressed in binary form as 
\begin{equation}
\begin{aligned}
    \varphi = & 0.\varphi_0 \varphi_1...\varphi_{t-1} \\
    = & \frac{1}{2}\varphi_0 + \frac{1}{2^2}\varphi_1 +...+ \frac{1}{2^t}\varphi_{t-1}.
\end{aligned}
\end{equation}
The quantum state in \eqref{qpe1} can be rewritten as 
\begin{equation}
\begin{aligned}
    |\psi_2\rangle = & \frac{1}{\sqrt{2^t}}(|0\rangle + e^{i2\pi0.\varphi_{t-1}}|1\rangle)(|0\rangle + e^{i2\pi0.\varphi_{t-2}\varphi_{t-1}}|1\rangle)...\\& (|0\rangle + e^{i2\pi0.\varphi_0\varphi_{t-1}}|1\rangle)|u\rangle.
\end{aligned}
\end{equation}
To convert the encoded phases into measurable binary information, the inverse quantum Fourier transform is applied to the phase register as 
\begin{equation}
    |\psi_3\rangle = \frac{1}{2^t}\sum_{m=0}^{2^t-1}\sum_{k=0}^{2^t-1}e^{i2\pi k(\varphi - m/2^t)}|m\rangle|u\rangle.
\end{equation}
The resulting quantum state is 
\begin{equation}
    |\psi_3\rangle = |2^t \tilde{\varphi}\rangle|u\rangle.
\end{equation}
By performing measurements on the qubits in the phase register, the phase, $\varphi$, can be obtained. 

These theoretical results suggest a potential exponential speedup compared to classical direct or iterative solvers. 
However, the practical realization of HHL requires deep quantum circuits, high-precision controlled operations, and long coherence times, rendering the algorithm highly sensitive to noise and difficult to scale on current noisy intermediate-scale quantum (NISQ) devices. 

In contrast to HHL, variational quantum algorithms (VQAs) are designed with near-term implementability in mind \cite{cerezo2021variational}, and are compatible with NISQ hardware \cite{qi2024variational}. 
VQAs approximate target quantum states using parameterized quantum circuits optimized through a hybrid quantum–classical loop. 
Representative examples include the variational quantum eigensolver \cite{kandala2017hardware}, and the variational quantum linear solver (VQLS) \cite{bravo2023variational}, which reformulate linear-systems solving as variational optimization problems inspired by the structure of HHL. 
Although VQLS does not inherit the asymptotic exponential speedup of HHL, it offers improved robustness against hardware noise and relaxes the need for deep quantum circuits and fault tolerance. 
For power system problems, where large, sparse matrices, measurement noise, and hardware imperfections are unavoidable, VQA-based linear solvers provide a more realistic pathway to exploring quantum acceleration in the field of optimization, dynamic analysis, and data-driven modeling tasks \cite{golestan2023quantum}. 

Overall, quantum linear solvers span a spectrum ranging from theoretically optimal but hardware-demanding algorithms, such as HHL, to practically executable but heuristically optimized approaches based on variational principles. 
Rather than guaranteeing immediate computational advantage, existing quantum linear solvers primarily indicate the potential for acceleration under favorable structural and hardware conditions. 
Understanding the tradeoffs among theoretical complexity, algorithmic robustness, and physical applicability remains a challenge in deploying quantum linear solvers for large-scale power system simulations. 

\subsection{Quadratic Unconstrained Binary Optimization}
\subsubsection{Mathematical Definition}
Quadratic unconstrained binary optimization (QUBO) provides a unified mathematical framework for representing a wide range of combinatorial optimization problems using binary decision variables \cite{glover2022quantum}. 
A standard QUBO problem is expressed as 
\begin{equation}
    \min _{\boldsymbol{x}\in\{0,1\}^N} \boldsymbol{x}^\transpose \boldsymbol{Q} \boldsymbol{x},
    \label{QUBO}
\end{equation}
where $\boldsymbol{x}$ denotes an $N$-dimensional binary vector and $\boldsymbol{Q} \in \mathbb{R}^{N \times N}$ is a symmetric coefficient matrix encoding interactions between variables.

QUBO model \eqref{QUBO} is mathematically equivalent to the Ising model \cite{brush1967history}, which is defined on spin variables $z_i \in \{ +1,-1 \}$ by the Hamiltonian as 
\begin{equation}
    H(z) = \sum_{i<j}^{} J_{ij}z_i z_j + \sum_{i=1}^{N} h_i z_i,
    \label{ising}
\end{equation}
where $z_i$ denotes the spin variable of site $i$, $h_i$ represents the local field acting on each spin, and $J_{ij}$ captures the pairwise interaction strength between spins $i$ and $j$. 
Note that $J_{ij} \neq 0$ if $z_i$ and $z_j$ are neighbors; otherwise, $J_{ij}=0$. 

The two models can be mapped via a linear transformation, i.e., $x_i=(1+z_i)/2$, which preserves the problem structure. 
This equivalence is fundamental because most quantum optimization algorithms and devices naturally operate in the Ising formulation \cite{glover2022quantum}, \cite{lucas2014ising}. 
The mapping between the QUBO model and the Ising model is derived as follows. 

The polynomial form of the QUBO model is
\begin{equation}
    \sum_{i=1}^N\sum_{j=1}^N q_{ij} x_i x_j + \sum_{i=1}^N q_i x_i + C_{QUBO},
\end{equation}
where $x_i$ is a binary variable, $q_{ij}$ and $q_i$ represent the coefficients of the quadratic and linear terms, $C_{QUBO}$ is a constant. 
Through the linear transformation, \eqref{ising} can be rewritten as 
\begin{equation}
\begin{aligned}
    H = & \sum_{i=1}^{N-1} \sum_{j=i+1}^N 4J_{ij} x_i x_j - \sum_{i=1}^{N-1}x_i \sum_{j=i+1}^N 2J_{ij} - \\ & \sum_{i=1}^{N-1} \sum_{j=i+1}^N 2J_{ij} x_j - \sum_{i=1}^N 2h_i x_i + C_{Ising}.
    \label{Hsing}
\end{aligned}
\end{equation}
Assuming $a_i = -\sum_{j=i+1}^N 2 J_{ij} + 2h_i$ and neglecting the constant terms, $C_{Ising}$, that do not influence the optimization, we obtain 
\begin{equation}
\begin{aligned}
    \sum_{i=1}^{N-1} \sum_{j=i+1}^N J_{ij} x_j = & J_{1N}x_N + J_{1,N-1} x_{N-1} + ... + J_{12}x_2 + \\ &  J_{2N}x_N+...+J_{23}x_3 + ... + \\ & J_{N-1,N}x_N \\ 
    = & \sum_{i=1}^N b_i x_i
\end{aligned}
\end{equation}
According to the equation, \eqref{Hsing} can be rewritten as 
\begin{equation}
\begin{aligned}
    H = & \sum_{i=1}^{N-1}\sum_{j=i+1}^N 4J_{ij} x_i x_j + \sum_{i=1}^{N-1}(a_i - 2b_i) x_i + \\ & (-2h_N-2b_N)x_N \\  = & \sum_{i=1}^{N-1}\sum_{j=i+1}^N 4J_{ij}x_i x_j + \sum_{i=1}^N c_i x_i.
    \label{Hsingg}
\end{aligned}
\end{equation}
By setting $q_{ij} = 2J_{ij}, q_i = c_i, C_{QUBO} = 0$, \eqref{Hsingg} can be transformed into 
\begin{equation}
    H = \sum_{i=1}^N\sum_{j=1}^Nq_{ij}x_ix_j + \sum_{i=1}^N q_i x_i.
\end{equation}
Note that $x_i$ is a binary variable that satisfies $x_i^2 = x_i$, the coefficients of the quadratic and linear terms can be combined, resulting in the equivalent formulation in \eqref{QUBO}. 

\subsubsection{Quantum Methods for Solving QUBO Problems}
QUBO has become the principal interface between classical optimization models and quantum hardware. 
Two major quantum approaches are widely used to solve QUBO or its equivalent Ising representation. 
The first is the Quantum Approximate Optimization Algorithm (QAOA) \cite{zhou2020quantum}, a variational algorithm that constructs a parameterized quantum circuit whose cost Hamiltonian directly corresponds to the QUBO matrix. 
The second approach uses specialized quantum hardware, including quantum annealers \cite{king2025beyond} and coherent Ising machines \cite{wang2013coherent}, which solve QUBO formulations by minimizing physical energy. 
Both position QUBO as the primary mathematical format for deploying optimization problems across diverse quantum platforms.

\subsubsection{Applications of QUBO Modeling in the Power System}
QUBO provides a natural modeling framework for a wide range of combinatorial optimization problems \cite{glover2022quantum}, \cite{lucas2014ising}, many of which are fundamental to power system planning and operation. 
Classical tasks such as unit commitment, network operation and reconfiguration, fault identification, and market clearing all involve binary or mixed-integer decisions that can be represented using binary variables and reformulated as a QUBO. 
Once these problems are encoded as a QUBO, they can be directly mapped to quantum optimization methods, enabling rapid search for high-quality solutions on quantum hardware. 
This capability positions QUBO as a key interface between classical power system optimization models and emerging quantum computational technologies \cite{morstyn2024opportunities}. 

\begin{figure}[htbp]
    \centering
    \includegraphics[width=0.97\linewidth]{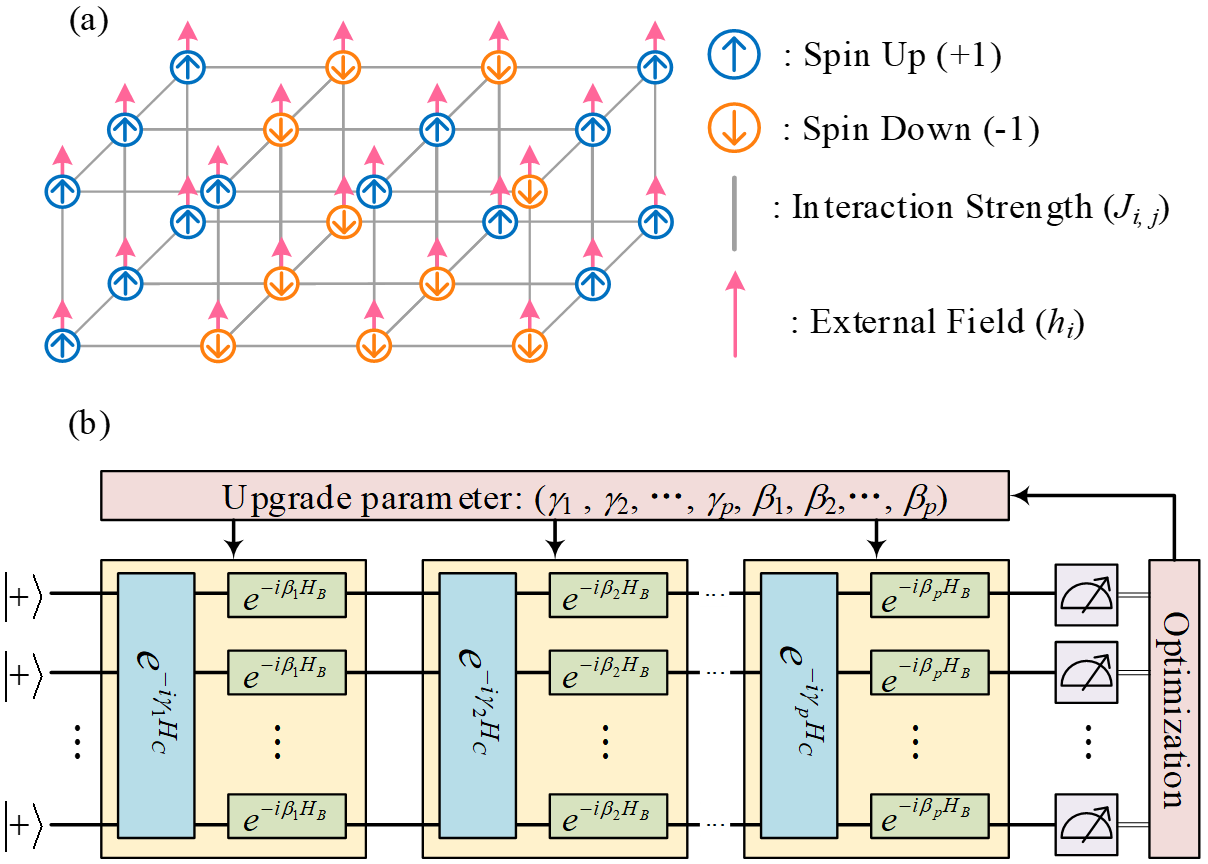}
    \caption{(a) Visualization of a 3D Ising model, and (b) Variational quantum circuit of the QAOA.}
    \label{fig2}
\end{figure}

With the presentation of the fundamentals of quantum computing, the following section reviews its application in power systems, with an emphasis on existing research. 

\section{Landscape For Quantum Applications in Power Systems}
The rapid evolution of modern power systems, driven by the large-scale integration of renewables and converter-dominated architectures, has significantly increased model complexity and computational demands \cite{golestan2023quantum,morstyn2024opportunities,cortes2018characterizing,chen2024distributed}. 
High-fidelity static and transient simulations, together with large-scale optimization tasks, are now required to support planning, operation, and control. 
However, the computational cost of conventional simulation and optimization methods continues to grow with system size, which poses challenges for real-time applications. 

In the face of these challenges, quantum computing has attracted growing interest as a potential computational paradigm for certain power system applications. 
By exploiting fundamentally different information-processing principles, quantum approaches can accelerate key computational components, such as large-scale linear solvers and combinatorial optimization. 
Existing studies have therefore begun to explore the role of quantum computing in power systems, with a particular focus on two directions: quantum-based simulation methods for power flow analysis and transient analysis and quantum-enabled optimization techniques for dispatch, network operation and market-related problems \cite{zhou2022quantum,ajagekar2022quantum,ajagekar2019quantum}. 
Essentially,  transient simulations can be formulated as differential equations-based forms, whereas optimization problems can be cast as integer programming or related models. Given the distinct mathematical forms, the quantum algorithms employed are fundamentally different. To this end, we next review the landscape of quantum computing in power systems from two aspects: simulation and optimization. 

\subsection{Quantum Computing Power System Simulation}
\begin{figure}[htbp]
    \centering
    \includegraphics[width=\linewidth]{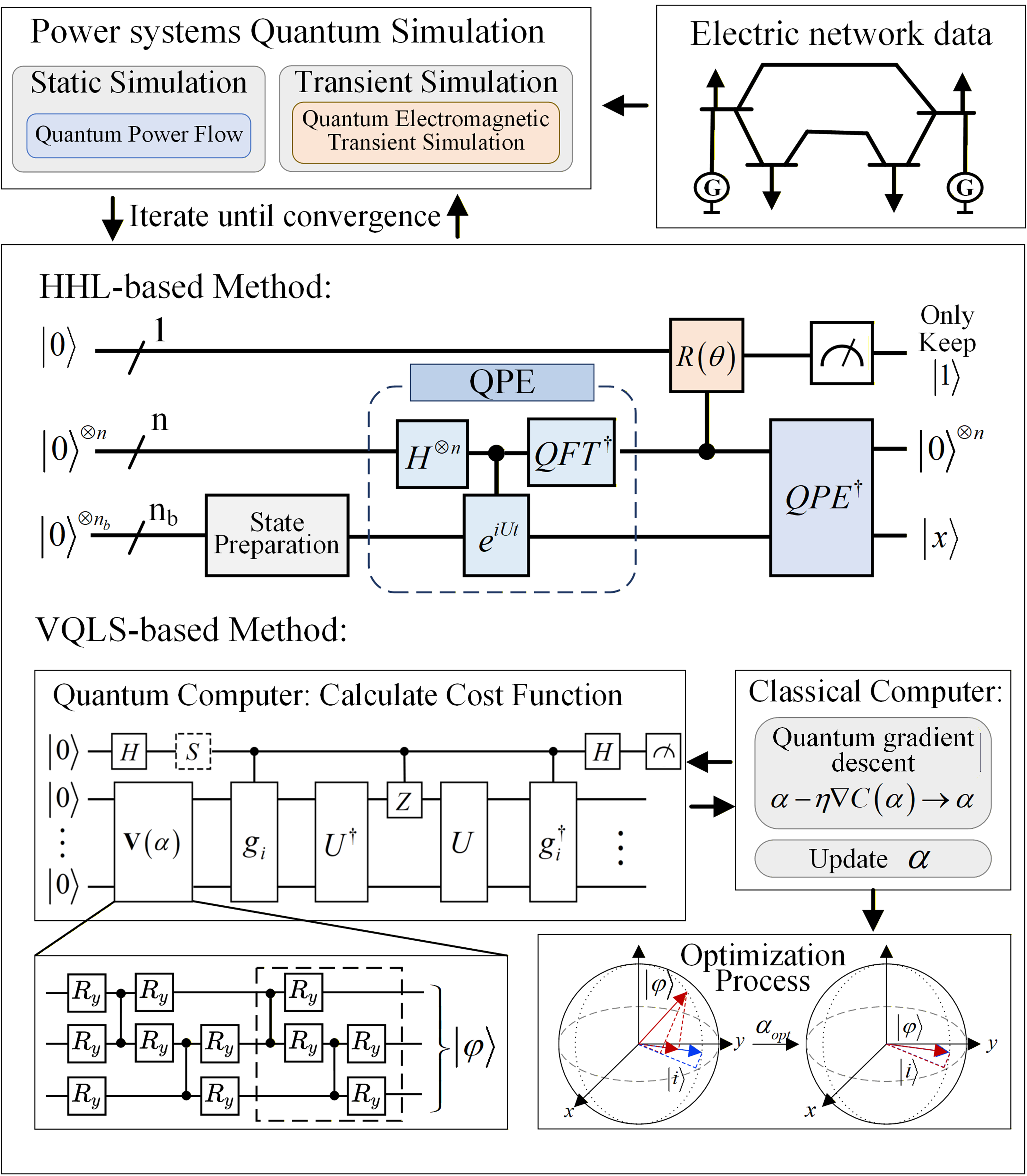}
    \caption{Algorithmic framework for quantum simulation of power systems. The figure illustrates the overall workflow of quantum power system simulation. Network parameters and system topology are first specified as inputs, based on which either steady-state or transient models are constructed. The resulting models are then solved using quantum algorithms, with representative quantum circuit implementations based on the HHL and VQLS solvers shown for illustration. The quantum solver iteratively evolves until convergence, or the designated simulation time is reached, after which the required electrical quantities are obtained as outputs.}
    \label{fig3}
\end{figure}

Quantum computing has been explored for steady-state and transient power system simulation. 
The overall algorithmic framework for the quantum simulation of power systems is illustrated in Fig. \ref{fig3}.

\begin{table*}[htbp]
    \centering
    \caption{Summary of Quantum Computing Applications in Power Systems}
    \begin{tabular}{cccccc}
    \toprule
    \textbf{Research Area} & \textbf{Detailed Field} & \textbf{References} & \textbf{Algorithm} & \textbf{Platform} & \textbf{Hybrid} \\
    \hline
    \multirow{4}{*}{Quantum Simulation} & \multirow{2}{*}{Power flow}
     & \cite{eskandarpour2020quantum, feng2021quantum, saevarsson2022quantum, gao2023solving} & HHL & Universal & $\times$ \\ 
     & & \cite{feng2023noise} & VQLS & Universal & \checkmark \\ \cline{2-6}
      & \multirow{2}{*}{EMT simulation}
     & \cite{zhou2021quantum} & HHL & Universal & $\times$ \\ 
     & & \cite{zhou2022noisy,lou2026matrix} & VQLS & Universal & \checkmark \\ \hline
     \multirow{15}{*}{Quantum Optimization} & \multirow{7}{*}{Power system dispatch} & \cite{nikmehr2022quantum, yang2025scalable, fu2025quantum, mahroo2023learning} & Q-ADMM & Dedicated & \checkmark \\
     & & \cite{paterakis2023hybrid, ling2025hybrid, ellinas2024hybrid, fu2023coordinated} & Benders decomposition & Dedicated & \checkmark \\
     & & \cite{feng2023noise} & Quantum Lagrangian relaxation & Dedicated & \checkmark \\
     & & \cite{zheng2024fast} & Grover search algorithm & Universal & $\times$ \\
     & & \cite{morstyn2022annealing, nikmehr2024quantum} & Quantum annealing & Dedicated & $\times$  \\
     & & \cite{amani2025optimal} & Quantum interior point method & Universal & $\times$ \\
     & & \cite{hafshejani2025quantum} & HHL, VQLS & Universal & $\times$ \\ \cline{2-6}
     & \multirow{5}{*}{Power network optimization} & \cite{kea2023leveraging} & Quantum annealing & Dedicated & $\times$ \\
     & & \cite{wang2022quantum, jing2022interoperation, jing2023data, colucci2023power, silva2022qubo, silva2023quantum, bi2023quantum} & Quantum annealing & Dedicated & \checkmark \\ 
     & & \cite{fei2024power} & Symmetric equivalent decomposition method & Dedicated & \checkmark \\
     & & \cite{kea2023leveraging} & QAOA & Universal & $\times$ \\
     & & \cite{li2025quantum} & Benders decomposition & Dedicated & \checkmark \\ \cline{2-6}
     & \multirow{3}{*}{Electricity market optimization} & \cite{ajagekar2024variational} & QAOA & Universal & $\times$ \\
     & & \cite{bucher2025grid} & Quantum annealing & Dedicated & \checkmark \\
     & & \cite{han2024quantum} & Benders decomposition & Dedicated & \checkmark \\
     \bottomrule     
    \end{tabular}
    \label{table1}
\end{table*}

\subsubsection{Quantum Static Simulation for Power Systems}
Static analysis, with power flow computation as its core, underpins most power system analysis tasks \cite{feng2022distributed}. 
Power flow analysis seeks to solve nonlinear algebraic equations that describe nodal power balance as determined by generators, loads, and network topology \cite{zhou2020reachable}. 
Classical solution techniques, including the Newton–Raphson method \cite{feng2020enhanced}, the Gauss-Seidel method \cite{9110755}, and the fast decoupled power flow \cite{stott2007fast}, all rely on repeated computing of large, sparse linear systems. 
Consequently, the efficiency of the underlying linear solver largely determines the scalability of power flow computation. 
As the complexity of conventional iterative methods scales polynomially with system size, their effectiveness in large-scale studies becomes increasingly limited. 
Quantum computing has thus emerged as a potential paradigm shift, offering the possibility of exponential acceleration for solving large, sparse linear systems. 

In 2020, Eskandarpour \emph{et al.} \cite{eskandarpour2020quantum} introduced the concept of Quantum Power Flow (QPF) by applying the HHL algorithm to DC power flow, marking the first formal integration of quantum algorithms into power system static analysis. 
In 2021, Feng \emph{et al.} \cite{feng2021quantum} extended this approach to AC power flow by incorporating fast decoupled power flow principles into the HHL framework. 
Leveraging the sparsity and near-Hermitian structure of Jacobian matrices, the work demonstrated the theoretical feasibility of AC QPF.

Subsequently, in 2022, Sævarsson \emph{et al.} \cite{saevarsson2022quantum} deployed HHL-based AC power flow on real quantum hardware, providing the first experimental validation of quantum power flow execution and revealing critical limitations related to noise, circuit depth, and hardware scale. 
To address these hardware constraints, Feng \emph{et al.} \cite{feng2023noise} proposed a noise-resilient QPF architecture based on VQLS and experimentally demonstrated scalable implementation on the IBMQ-Hanoi processor for the IEEE 118-bus system, showing that shallow variational circuits enable practical execution on NISQ hardware. 
Most recently, Gao \emph{et al.} \cite{gao2023solving} proposed a hybrid multiple-phase estimation algorithm that significantly reduces qubit requirements and circuit depth, paving the way for resource-efficient quantum power flow solvers. 

\subsubsection{Quantum Transient Simulation for Power Systems}
The increasing penetration of power electronic converters and distributed energy resources continue to reduce system inertia, threatening system stability \cite{joint2021odessa}. 
To accurately characterize the high-frequency dynamics and switching behavior of converter-dominated systems, electromagnetic transient (EMT) simulation has become essential \cite{watson2003power}. 
While classical EMT programs provide detailed multi-rate waveform modeling \cite{zhang2010shifted}, their computational complexity scales to $O(N^3)$, posing prohibitive challenges for real-time analysis or modeling of large converter-based power systems \cite{zhang2017key}. 

Quantum computing presents a fundamentally different approach to addressing this challenge \cite{hidary2021quantum}. 
In 2021, Zhou \emph{et al.} \cite{zhou2021quantum} introduced the first quantum electromagnetic transient program (QEMTP) based on HHL, demonstrating a hybrid quantum–classical framework to model EMT behavior. 
Through analysis, the comparison between the HHL-based QEMTP and the optimal classical algorithm is presented in Table \ref{tablehhl}. 
It shows that, in theory, the HHL-based QEMTP achieves an exponential reduction in time complexity. 
Note that the quantum advantage offered by the algorithm is established under idealized assumptions and favorable conditions. 
In practice, the performance of the algorithm is influenced by several factors, including matrix sparsity, the condition number, input quantum state preparation, hardware noise, and measurement readout. 
In particular, extracting solution information from the output quantum state may introduce additional computational overhead, diminishing the theoretical speedup. 

\begin{table}[htbp]
    \centering
    \caption{Comparison between the HHL-Based QEMTP and the \\ Optimal Classical Algorithm}
    \begin{tabular}{cc}
    \toprule
    \textbf{Algorithm} & \textbf{Time Complexity}  \\
    \hline
    Optimal classical algorithm & $\mathcal{O}(N)$ \\
    HHL-based QEMTP & $\mathcal{O}(\log(N))$ \\
    \bottomrule     
    \end{tabular}
    \label{tablehhl}
\end{table}

However, considering current NISQ, the fault-tolerant hardware required by the HHL algorithm is unlikely to be realized in the foreseeable future \cite{duan2020survey,preskill2018quantum}. 
Under this practical limitation, Zhou \emph{et al.} \cite{zhou2022noisy} further proposed a VQLS-based NISQ-QEMTP framework in 2023, extending QEMTP from idealized quantum simulators to real quantum processors. 
The approach employs shallow parameterized circuits and batch execution strategies to reduce redundant circuit evaluations. 
Experimental results on IBM quantum hardware demonstrated successful QEMTP on systems with up to 906 nodes, showing improved noise tolerance, convergence stability, and scalability compared with prior HHL-based approaches. 
The comparison between HHL and VQLS is presented in Table \ref{tablehhlvqls}.

\begin{table}[htbp]
    \centering
    \caption{Comparison between HHL and VQLS}
    \begin{tabular}{cccc}
    \toprule
    \textbf{Algorithm} & \textbf{Complexity} & \textbf{\makecell{Required Qubits}} & \textbf{\makecell{Execution \\ on NISQ}}  \\
    \hline
    HHL & $\mathcal{O}(\log(N))$ & $\mathcal{O}(N+\log_2(N)+1)$ & Not suitable \\
    VQLS & \makecell{Higher than \\ HHL} & $\mathcal{O}(\log_2(N))$ & Suitable \\
    \bottomrule     
    \end{tabular}
    \label{tablehhlvqls}
\end{table}

Despite this progress, variational methods still struggle with highly dynamic networks containing switching-converter components, where the network topology evolves frequently and circuit parallelism increases sharply. 
To address this challenge, Lou \emph{et al.} \cite{lou2026matrix} introduced the first QEMTP framework based on a fixed-admittance switch model. 
By reducing repeated matrix updates and leveraging matrix low-qubit casting, the framework has the potential to improve preprocessing efficiency and reduce parallel circuit requirements. 

\subsection{Quantum Computing in Power System Optimization}
\subsubsection{Applications in Power System Dispatch}
The unit commitment (UC) problem, a core optimization task in power system dispatch, is typically expressed as a mixed-integer programming (MIP) formulation. 
Its nonconvex and combinatorial nature has driven growing interest in quantum optimization methods to enhance computation efficiency. 

Paterakis \emph{et al.} \cite{paterakis2023hybrid} proposed a classical-quantum hybrid multi-cut Benders decomposition algorithm that explicitly considers the temporal coupling characteristics of UC. 
Building on this, Ling \emph{et al.} \cite{ling2025hybrid} summarized a general hybrid Benders decomposition-based framework for UC and verified its feasibility on real quantum hardware. 
To overcome the limitation that Benders decomposition can only be applied to MILP models, Feng \emph{et al.} \cite{feng2022novel} introduced a distributed quantum Lagrangian relaxation algorithm within the QAOA framework, enabling broader UC formulations. 

To further enhance the scalability of decomposition and the convergence speed, the quantum alternating direction method of multipliers (Q-ADMM) was proposed. 
The comparative analysis of the complexity between Q-ADMM and its classical counterpart is provided in Table \ref{tableqadmm}. 
Note that the analysis presented in this table assumes idealized conditions. 
The practical efficiency of Q-ADMM is affected by factors such as the cost of block encoding construction and the overhead associated with quantum-classical information exchange. 
Nikmehr \emph{et al.} \cite{nikmehr2022quantum} formulated the UC problem in a quantum setting and designed a Q-ADMM-based distributed coordination mechanism to alleviate hardware qubit constraints, which was experimentally validated on a physical quantum processor. 
Experimental results illustrate that, in a microgrid with three distributed resources, the operational costs obtained by the quantum and classical algorithms exhibit a small difference. 
In addition, the quantum algorithm requires $3$ iterations to converge, compared with $8$ iterations for the classical algorithm, exhibiting better convergence performance. 
Subsequently, Yang \emph{et al.} \cite{yang2025scalable} introduced a consensus-constraint encoding and a divide-and-conquer strategy to further reduce qubit requirements. 
Extending these efforts, Fu \emph{et al.} \cite{fu2025quantum} developed a two-stage robust UC model that considers extreme-weather resilience constraints and solved it using a combination of Q-ADMM and column-generation algorithms. 
Distinct from hybrid reformulations, Zheng \emph{et al.} \cite{zheng2024fast} employed the Grover search algorithm to achieve quantum-native optimal-solution retrieval for the UC problem.

\begin{table}[htbp]
    \renewcommand{\arraystretch}{1.2}
    \centering
    \caption{Comparison between Q-ADMM and Classical ADMM}
    \begin{tabular}{cc}
    \toprule
    \textbf{Algorithm} & \textbf{Complexity}  \\
    \hline
    Classical ADMM & $\mathcal{O}(N^6)$ \\
    Q-ADMM & $\mathcal{O}(N^2)$ \\
    \bottomrule     
    \end{tabular}
    \label{tableqadmm}
\end{table}

Optimal power flow (OPF) problems extend UC by incorporating network constraints. 
Although classical OPF is formulated as a continuous optimization problem, recent studies have explored quantum-oriented reformulations to enable the application of quantum optimization methods \cite{kaseb2024power}. 
Ellinas \emph{et al.} \cite{ellinas2024hybrid} and Mahroo \emph{et al.} \cite{mahroo2023learning} combined Benders decomposition and Q-ADMM frameworks to design a hybrid decomposition approach for general OPF, validated on quantum hardware.
Furthermore, Morstyn \cite{morstyn2022annealing} proposed the concept of combinatorial OPF, reformulated it as a QUBO problem to incorporate distributed energy resources. 
To address noise sensitivity in real quantum hardware, Amani and Kargarian \cite{amani2025optimal} presented a noise-resilient quantum interior-point method for DC-OPF. 
Finally, Hafshejani \emph{et al.} \cite{hafshejani2025quantum} demonstrated the superior performance of HHL and VQLS algorithms compared with classical solvers for OPF optimization. 

Building upon UC and OPF formulations, researchers have begun exploring more complex dispatch problems. 
For instance, in time-critical resilience scheduling, Fu \emph{et al.} \cite{fu2023coordinated} and Nikmehr \emph{et al.} \cite{nikmehr2024quantum} adopted the QAOA-based optimization models for critical load restoration in distribution networks, verifying the advantages of quantum computation for this topic. 

\subsubsection{Applications in Power Network Optimization}
Power network partitioning, a typical NP-hard combinatorial problem, has recently emerged as a key area showcasing the potential superiority of quantum computing over classical methods \cite{bucher2024evaluating}. 
Wang \emph{et al.} \cite{wang2022quantum} pioneered the application of quantum computing to power network partitioning and confirmed its feasibility. 
Building upon this, Jing \emph{et al.} \cite{jing2022interoperation, jing2023data} formulated the problem as a weighted max-cut model and implemented quantum solutions using QAOA on real quantum hardware. 
Subsequent work focused on refining the application scenarios and improving model fidelity; Colucci \emph{et al.} \cite{colucci2023power} proposed a network partitioning model for power-surplus optimization. 

Apart from network partitioning, related topics such as network reconfiguration, fault localization, and path optimization have also attracted attention \cite{liu2026alleviating}. 
For network reconfiguration, Silva \emph{et al.} \cite{silva2022qubo, silva2023quantum} developed distribution-network models minimizing power loss and designed hybrid quantum-classical solvers. 
In fault localization, Bi \emph{et al.} \cite{bi2023quantum} formulated the fault diagnosis problem as a QUBO model solved via an improved quantum annealing algorithm, while Fei \emph{et al.} \cite{fei2024power} optimized qubit usage through multi-Z rotation symmetry decomposition for detecting low-probability fault events. 
For path optimization, Kea \emph{et al.} \cite{kea2023leveraging} proposed a QAOA-based EV charging-route optimization, and Li \emph{et al.} \cite{li2025quantum} presented a hybrid QUBO model for coordinated hydrogen-electric distribution operation solved by a Benders decomposition-based hybrid framework. 

\subsubsection{Applications in Electricity Market Optimization}
Optimization in the electricity market imposes extremely high requirements on computing efficiency, since market clearing, bidding strategies, and resource allocation are computed within very short operational intervals to ensure fair and reliable market operation. 
Quantum computing has been applied in a range of scenarios within the field. 
Existing studies span demand response \cite{ajagekar2024variational}, peer-to-peer trading \cite{bucher2025grid}, and economic dispatch \cite{han2024quantum}. 
These studies employ various quantum hardware, including general-purpose quantum computers \cite{ajagekar2024variational} and quantum annealers \cite{bucher2025grid,han2024quantum}. 
This diversity highlights a trend of multi-hardware and multi-scenario co-development in electricity market research. 
Relevant summaries are provided in Table \ref{table1}.

\subsection{Current Issues}
Despite promising progress, quantum simulation and optimization for power systems remain hindered by four fundamental barriers: 

\subsubsection{Algorithmic Trainability}
Variational quantum circuits exhibit sensitivity to ill-conditioned matrices \cite{saad2003iterative}, which is typical of EMT models \cite{wang2015emtp}. 
This poses a significant challenge for the training of variational quantum circuits \cite{hosaka2023preconditioning}. 

\subsubsection{Qubit Count Limitations}
The number of reliable, high-fidelity qubits available on current devices remains far below what is required for realistic power system applications. 
This limits the size of embeddable optimization and simulation problems, thereby preventing large-scale instances from being executed on existing hardware. 

\subsubsection{Resource and Circuit Complexity}
Both quantum simulation and optimization workflows often rely on repeated circuit evaluations and parameter updates. 
Current architectures still require a large number of parallel circuit evaluations, and in many cases, available quantum hardware cannot support increasingly deep circuits \cite{turati2024empirical}. 

\subsubsection{Quantum Error}
The dominant sources of error in quantum computers are imperfect operations, unwanted coupling to the environment causing decoherence, and noise in state preparation and measurement, all of which accumulate with circuit depth and system size. 
Quantum error mitigation and correction address noise at different conceptual layers and with very different assumptions about hardware maturity. 
Error mitigation acknowledges that errors occur on today’s NISQ hardware but seeks to reduce their impact on final observables through algorithmic and statistical post-processing, such as zero-noise extrapolation, probabilistic error cancellation, symmetry verification, or measurement-error mitigation. This means that individual circuit runs remain noisy, but ensemble averages are debiased at the cost of increased sampling overhead and without any fault-tolerance guarantees \cite{cai2023quantum}. 
Error correction, is an active, fault-tolerant framework that encodes logical qubits into many physical qubits, repeatedly measures error syndromes without collapsing the logical state, and applies conditional recovery operations so that errors are detected and corrected during computation, enabling arbitrarily long computations provided physical error rates are below a threshold, but at the expense of substantial qubit and circuit overhead and strict architectural requirements \cite{devitt2013quantum}. 

\subsubsection{Hardware Reliability}
Limited qubit connectivity, short coherence time, and hardware noise lead to solution drift and instability, preventing accurate simulation of large-scale systems \cite{chen2024enabling}. 
Furthermore, the HHL algorithm and the VQLS assume that the quantum error signals are stationary and linearly dependent. However, by analyzing real data obtained from an IBM quantum computer, Saghafi and Mili \cite{saghafi2025predictive} demonstrated that quantum output data generated by quantum hardware exhibit long-memory behavior and can be predicted based on their historical observations. Furthermore, Saghafi \emph{et al.} \cite{saghafi2025multifractality} showed that these signals exhibit significant systematic biases and multifractal characteristics, indicating complex scaling behavior and nonlinear dependencies.
The multifractal properties of the signal were revealed through the singularity spectrum, estimated using the wavelet leader method. 
Recall that the singularity spectrum is a plot of a function $D(h)$, where $D(\cdot)$ is the Hausdorff (fractal) dimension and $h$ is the Holder exponent, which measures the degree of singularity. 
Consequently, to achieve practical quantum simulation and optimization for power systems, breakthroughs are required on multiple fronts, such as algorithmic robustness, quantum hardware scalability, and noise-mitigation techniques \cite{preskill2018quantum}. 

Currently, research on quantum computing in power systems is extensive, focusing primarily on two aspects: simulation and optimization, both of which involve solving mathematical models. In addition, quantum computing can process data directly and is often combined with AI techniques, as elaborated in the next section.

\section{Roadmap for AI-Enabled Evolution}
Existing studies have demonstrated the potential of quantum computing for addressing a wide range of power system problems. 
Meanwhile, with the rapid advancement of AI, the integration of quantum computing and AI is expected to broaden the scope of quantum technologies in power system applications. 

\begin{figure}[htbp]
    \centering
    \includegraphics[width=\linewidth]{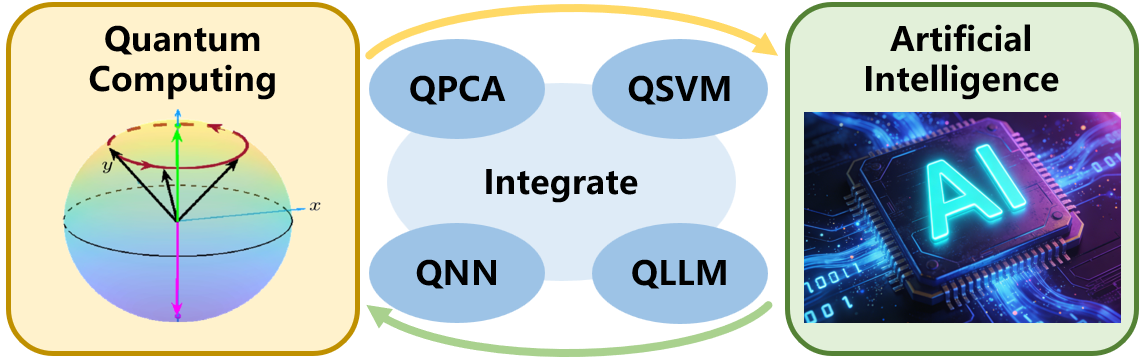}
    \caption{Integration of quantum computing and AI.}
    \label{fig4}
\end{figure}

In many tasks, classical AI has reached a relatively mature stage. 
However, as the complexity of power system models and problems continues to increase, classical AI faces challenges associated with substantial computational and energy costs. 
In this context, the motivation for incorporating quantum computing is to explore whether it can provide complementary advantages in specific computing tasks. 
Currently, a commonly adopted approach is to employ a quantum-classical hybrid framework, as shown in Fig. \ref{fig4}. 
Specifically, classical AI provides the overall framework, while quantum computing handles specific computational modules within it. 
This hybrid paradigm not only leverages the maturity and established capabilities of AI but also provides a pathway for exploring the potential of quantum computing. 
Such integration may help alleviate the bottlenecks encountered by classical AI. 
However, their application in power systems is currently very limited. 
To this end, this section outlines a roadmap for quantum computing with AI to empower power systems, summarized in Table \ref{table2}.

\begin{table*}[htbp]
\caption{Quantum AI-Enabled Evolution in Power Systems}
\centering
\renewcommand{\arraystretch}{1.1}
\scalebox{0.89}{
\begin{tabular}{cccccccccc}
\toprule
\textbf{Field} & \textbf{References} & \textbf{Algorithm} & \textbf{\makecell{Evidence \\ Type}} & \textbf{Task} & \textbf{\makecell{Problem \\ Scale}} & \textbf{\makecell{Input \\ Dimensionality}} & \textbf{\makecell{Reported \\ Quantum \\ Resources}} & \textbf{\makecell{Principal \\ Limitations}} & \textbf{\makecell{Potential \\ Power \\ System \\ Applications}} \\
\hline
\multirow{8}{*}{\makecell{Quantum \\ feature \\ learning}} & \cite{lloyd2014quantum} & \multirow{3}{*}{\makecell{Quantum principal \\ component analysis}} & Theoretical & -- & -- & -- & -- & \makecell{State \\ preparation} & \multirow{8}{*}{\makecell{Dimensionality \\ reduction}} \\ 
& \cite{xin2021experimental} & & Experimental & \makecell{Face \\recognition} & \makecell{Small-scale \\ image dataset} & $100$ & $4$-qubit & \makecell{Limited \\ scale} \\
& \cite{rebentrost2014quantum} & \multirow{4}{*}{\makecell{Quantum support \\ vector machine}} & Theoretical & -- & -- & -- & -- & \makecell{Data \\ loading \\ bottleneck} \\ 
& \cite{ding2024zc} & & Experimental & \makecell{Signal \\ classification} & $1000$ samples & $28$ & -- & \makecell{Limited \\ feature \\ dimension} \\ \hline
\multirow{8}{*}{\makecell{Quantum \\ predictive \\ learning}} & \cite{zhou2022noise} & \multirow{8}{*}{\makecell{Quantum \\ supervised \\ learning}} & Experimental & \makecell{(\textbf{Power}) \\ Stability \\ assessment} & \makecell{IEEE $300$-bus \\ system} & $1699$ & $2$-qubit & \makecell{Limited \\ scalability} & \multirow{8}{*}{\makecell{Dynamic \\ security \\ assessment \\ \\ Fault \\ diagnosis}}  \\ 
& \cite{sebastianelli2021circuit} & & Experimental & \makecell{Image \\ classification} & $27000$ samples & -- & $4$-qubit & \makecell{Quantum \\ noise} \\ 
& \cite{wang2024shallow} & & Experimental & \makecell{Image \\ classification} & \makecell{MNIST \\ dataset} & $784$ & $4$-qubit & \makecell{Limited \\ scalability} \\
& \cite{gong2024quantum} & & Experimental & \makecell{Image \\ classification} & \makecell{MNIST \\ dataset} & $784$ & $8$-qubit & \makecell{Limited \\ qubit \\ scale} \\ \hline
\multirow{5}{*}{\makecell{Quantum \\ generative \\ learning}} & \cite{zhu2022generative} & \multirow{5}{*}{\makecell{Quantum generative \\ adversarial networks}} & Experimental & \makecell{Financial \\ modeling} & \makecell{Joint \\ distribution}  & $2$ & $6$-qubit & \makecell{Small \\ dimension} & \multirow{5}{*}{\makecell{Scenario \\ generation}} \\ 
& \cite{tsang2023hybrid} & & Experimental & \makecell{Image \\ generation} & \makecell{MNIST \\ dataset} & $784$ & $8$-qubit & \makecell{Barren \\ plateau}  \\
& \cite{horowitz2022quantum} &  & Experimental & \makecell{Time-series \\ generation} & \makecell{Small \\ scale} & $2$ & $5$-qubit & \makecell{Limited \\ resources} \\ \hline
\multirow{5}{*}{\makecell{Quantum \\ probabilistic \\ inference}} & \cite{tucci1997quantum} & \multirow{5}{*}{Quantum Bayesian} & Theoretical & -- & -- & -- & -- & \makecell{Theory \\ oriented} & \multirow{5}{*}{\makecell{Uncertainty \\ analysis \\ \\ Parameter \\ inference}} \\ 
& \cite{low2014quantum} &  & Theoretical & -- & -- & -- & -- & \makecell{State \\ preparation} \\
& \cite{berner2021quantum} &  & Experimental & \makecell{Linear \\ regression} & \makecell{Small \\ scale} & $2$ & $5$-qubit & \makecell{Quantum \\ noise} \\ \hline
\multirow{3}{*}{\makecell{Quantum \\ reasoning}} & \cite{zhang2025llm} & LLM-QUBO & Experimental & Optimization & $100 \times 1000$ & -- & -- & \makecell{Limited \\ scalability} & \multirow{3}{*}{\makecell{Intelligent \\ analysis}} \\ 
& \cite{liu2025quantum} & \makecell{Quantum relational \\ knowledge distillation} & Experimental & \makecell{Image \\ classification} & \makecell{MNIST \\ dataset} & -- & $4$-qubit & \makecell{Hardware \\ noise}  \\
\bottomrule
\end{tabular}}
\label{table2}
\end{table*}

\subsection{Quantum Feature Learning}
One of the core tasks of AI is to uncover the intrinsic structures and patterns underlying massive datasets. 
For example, in classical approaches, principal component analysis reduces dimensionality through eigen-decomposition of the covariance matrix \cite{abdi2010principal}, while a support vector machine can identify optimal classification hyperplanes \cite{suthaharan2016support}. 
However, as data dimensionality and complexity grow exponentially, classical computers face severe computational bottlenecks when handling high-dimensional geometric data structures. 

To address these challenges, quantum computing offers a transformative pathway. 
Lloyd \emph{et al.} \cite{lloyd2014quantum} pioneered the theoretical framework of quantum principal component analysis (QPCA). 
This algorithm maps classical data into quantum state density matrices via quantum-state preparation techniques, then performs matrix exponentiation via Hamiltonian simulation. 
Subsequently, the quantum Fourier transformation is applied to obtain the eigenvectors corresponding to the principal components, along with the associated quantum state information. 
However, implementing this algorithm requires substantial computational resources, posing significant challenges for experimental verification. To this end, based on parameterized quantum circuits, Xin \emph{et al.} \cite{xin2021experimental} proposed an experimentally friendly QPCA algorithm, which was subsequently implemented for face recognition. 

Furthermore, quantum support vector machines (QSVMs) enhance dimensionality reduction in feature space mapping for data classification problems. 
Rebentrost \emph{et al.} \cite{rebentrost2014quantum} achieved a breakthrough in kernel function computation by mapping data into a high-dimensional quantum feature space, employing quantum parallelism to compute the kernel matrix. 
Leveraging this technology, Ding \emph{et al.} \cite{ding2024zc} analyzed particle collision experimental data, identifying and reconstructing the invariant mass distribution of a specified resonance state. 
This work successfully validated the feasibility of QSVM for high-dimensional data classification. 
The complexity comparison of QPCA, QSVM, and their classical counterparts is provided in Table \ref{tableqpca}. 
It is worth emphasizing that the complexity analysis assumes idealized conditions. 
In practice, the performance of QPCA depends on whether the input data can be efficiently represented as a density matrix, as well as on the implementation cost associated with the required quantum state evolution.
In addition, the practical performance of QSVM is sensitive to the quantum feature map and the structure of the input data, while reducing statistical errors requires a large number of measurement shots, which may introduce substantial computational overhead.

\begin{table}[htbp]
\caption{Comparison of QPCA, QSVM, and Their Classical Counterparts}
\centering
\renewcommand{\arraystretch}{1.2}
\begin{tabular}{ccccc}
\toprule
\textbf{Algorithm} & \textbf{Complexity}  \\
\hline
Classical PCA & $\mathcal{O}(N^2)$ \\
QPCA & $\mathcal{O}[(\log N)^2]$ \\ \hline
Classical SVM & $\mathcal{O}(N^3)$ \\
QSVM & $\mathcal{O}(\log N)$ \\
\bottomrule
\end{tabular}
\label{tableqpca}
\end{table}

For high-dimensional, complicated power system data, classical approaches are encountering bottlenecks in computing efficiency. 
In contrast, QPCA and QSVM theoretically offer an exponential decrease in computational complexity. This quantum-enabled AI approach facilitates efficient processing of high-dimensional, massive power system data, with further applications in feature selection, bad-data identification, and classification. 

At present, studies on quantum feature learning are mainly supported by theoretical analysis and classical simulation, while small-scale experiments have also been conducted on real quantum hardware. 
In the near term, quantum kernels and small-scale QSVMs can be implemented on NISQ devices to process low-dimensional data. 
In the medium term, increased qubit numbers and improved gate fidelity may allow larger feature spaces and datasets to be considered. 
In the long term, fault-tolerant quantum devices may enable large-scale QPCA and QSVM, allowing higher-dimensional data from large interconnected power systems to be processed. 

\subsection{Quantum Predictive Learning}
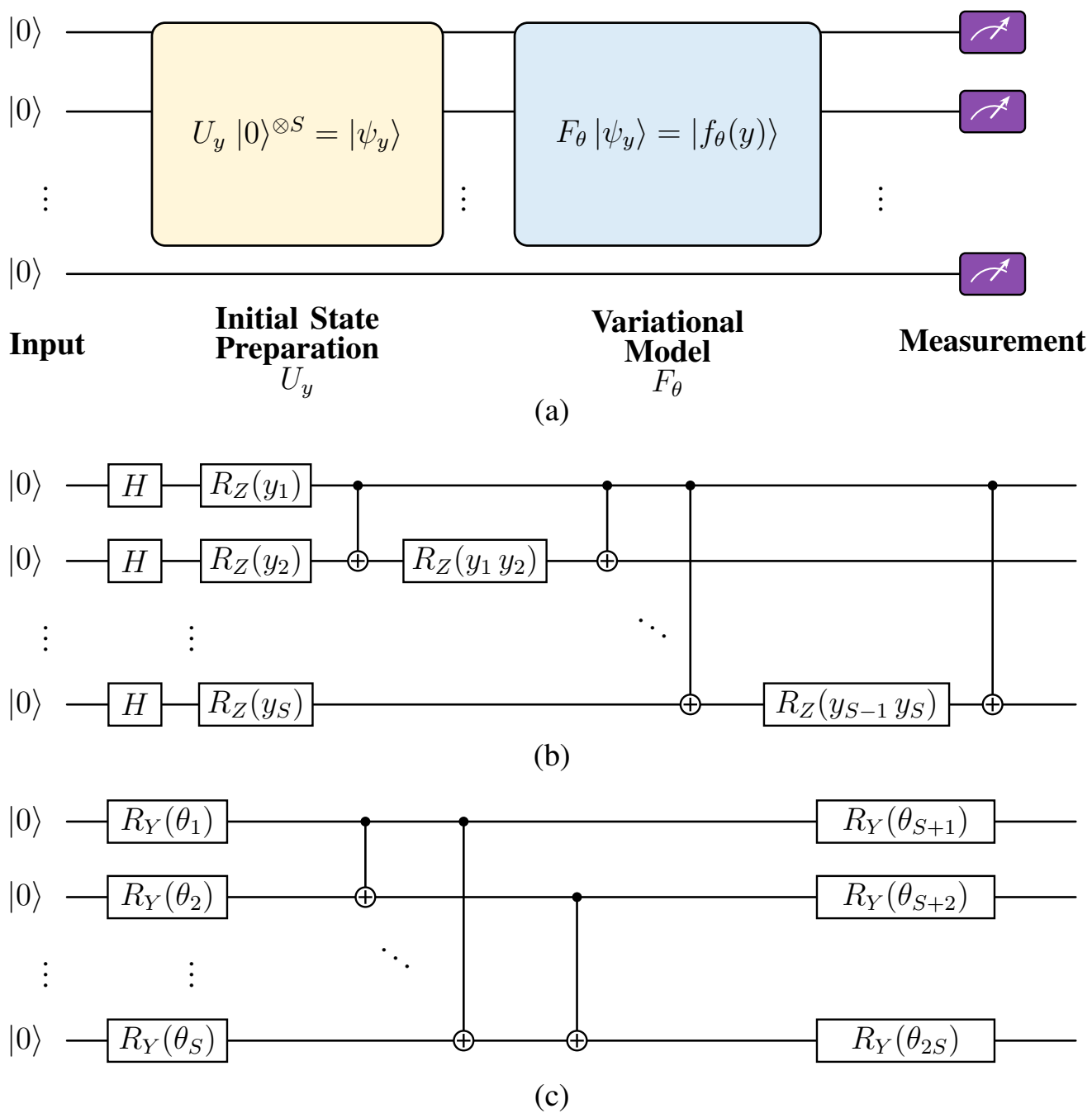
\begin{figure}[htbp]
    \centering
    \resizebox{\columnwidth}{!}{\begin{tikzpicture}[
    x=1cm,y=1cm,
    font=\fontsize{12}{12}\selectfont,
    wire/.style={line width=0.82pt, line cap=round},
    gate/.style={draw, fill=white, line width=0.75pt,
        minimum height=0.56cm, inner xsep=3pt, inner ysep=0pt,
        align=center},
    bigbox/.style={draw, rounded corners=5pt, line width=0.85pt,
        minimum height=2.95cm, align=center},
    labeltext/.style={font=\bfseries\fontsize{12}{12}\selectfont, align=center}
]

\newcommand{\ketzero}[2]{\node[anchor=east] at (#1,#2) {$\vert 0\rangle$};}
\newcommand{\ctrl}[2]{\fill (#1,#2) circle[radius=0.065cm];}
\newcommand{\targ}[2]{%
    \draw[fill=white,line width=0.75pt] (#1,#2) circle[radius=0.13cm];
    \draw[line width=0.75pt] (#1-0.09,#2) -- (#1+0.09,#2);
    \draw[line width=0.75pt] (#1,#2-0.09) -- (#1,#2+0.09);
}
\newcommand{\gatebox}[4]{\node[gate, minimum width=#3] at (#1,#2) {#4};}
\newcommand{\meas}[2]{%
    \node[draw, rounded corners=2pt, fill=measpurple, line width=0.75pt,
        minimum width=0.86cm, minimum height=0.55cm, inner sep=0pt] at (#1,#2) {};
    \begin{scope}[shift={(#1,#2)}, line cap=round, line join=round]
        \draw[white,line width=0.65pt]
            (-0.28,-0.10) .. controls (-0.17,0.07) and (0.03,0.12) .. (0.18,0.03);
        \draw[white,line width=0.85pt,-{Stealth[length=1.45mm,width=1.15mm]}]
            (-0.02,-0.08) -- (0.23,0.21);
    \end{scope}
}

\def\xa{0.70}
\def\xb{14.05}
\foreach \y in {13.70,12.65,10.50}{\draw[wire] (\xa,\y) -- (12.52,\y);}
\ketzero{0.52}{13.70}
\ketzero{0.52}{12.65}
\ketzero{0.52}{10.50}
\node at (0.42,11.60) {$\vdots$};

\node[bigbox, fill=initfill, minimum width=3.85cm] at (3.75,12.35)
    {$U_y\;\vert 0\rangle^{\otimes S}=\vert\psi_y\rangle$};
\node[bigbox, fill=varfill, minimum width=4.05cm] at (8.65,12.35)
    {$F_\theta\,\vert\psi_y\rangle=\vert f_\theta(y)\rangle$};

\node at (5.95,11.60) {$\vdots$};
\node at (11.47,11.60) {$\vdots$};
\meas{12.95}{13.70}
\meas{12.95}{12.65}
\meas{12.95}{10.50}

\node[labeltext] at (0.44,9.54) {Input};
\node[labeltext] at (3.75,9.68) {Initial State\\[-1pt] Preparation};
\node at (3.75,9.06) {$U_y$};
\node[labeltext] at (8.65,9.68) {Variational\\[-1pt] Model};
\node at (8.65,9.06) {$F_\theta$};
\node[labeltext] at (12.95,9.62) {Measurement};
\node at (7.12,8.67) {(a)};

\def\btop{7.70}
\def\bmid{6.70}
\def\bbot{4.80}
\foreach \y in {\btop,\bmid,\bbot}{\draw[wire] (0.70,\y) -- (14.05,\y);}
\ketzero{0.52}{\btop}
\ketzero{0.52}{\bmid}
\ketzero{0.52}{\bbot}
\node at (0.42,5.78) {$\vdots$};
\node at (2.35,5.78) {$\vdots$};

\gatebox{1.60}{\btop}{0.70cm}{$H$}
\gatebox{1.60}{\bmid}{0.70cm}{$H$}
\gatebox{1.60}{\bbot}{0.70cm}{$H$}
\gatebox{3.20}{\btop}{1.38cm}{$R_Z(y_1)$}
\gatebox{3.20}{\bmid}{1.38cm}{$R_Z(y_2)$}
\gatebox{3.20}{\bbot}{1.38cm}{$R_Z(y_S)$}

\draw[wire] (4.55,\btop) -- (4.55,\bmid); \ctrl{4.55}{\btop} \targ{4.55}{\bmid}
\gatebox{6.10}{\bmid}{1.85cm}{$R_Z(y_1\,y_2)$}
\draw[wire] (7.85,\btop) -- (7.85,\bmid); \ctrl{7.85}{\btop} \targ{7.85}{\bmid}

\node at (8.45,5.92) {$\ddots$};

\draw[wire] (8.95,\btop) -- (8.95,\bbot); \ctrl{8.95}{\btop} \targ{8.95}{\bbot}
\gatebox{11.15}{\bbot}{2.45cm}{$R_Z(y_{S-1}\,y_S)$}
\draw[wire] (12.95,\btop) -- (12.95,\bbot); \ctrl{12.95}{\btop} \targ{12.95}{\bbot}
\node at (7.12,4.10) {(b)};

\def\ctop{3.25}
\def\cmid{2.25}
\def\cbot{0.35}
\foreach \y in {\ctop,\cmid,\cbot}{\draw[wire] (0.70,\y) -- (14.05,\y);}
\ketzero{0.52}{\ctop}
\ketzero{0.52}{\cmid}
\ketzero{0.52}{\cbot}
\node at (0.42,1.33) {$\vdots$};
\node at (2.35,1.33) {$\vdots$};

\gatebox{2.03}{\ctop}{1.58cm}{$R_Y(\theta_1)$}
\gatebox{2.03}{\cmid}{1.58cm}{$R_Y(\theta_2)$}
\gatebox{2.03}{\cbot}{1.58cm}{$R_Y(\theta_S)$}

\draw[wire] (4.65,\ctop) -- (4.65,\cmid); \ctrl{4.65}{\ctop} \targ{4.65}{\cmid}
\draw[wire] (5.95,\ctop) -- (5.95,\cbot); \ctrl{5.95}{\ctop} \targ{5.95}{\cbot}
\node at (5.05,1.55) {$\ddots$};
\draw[wire] (7.45,\cmid) -- (7.45,\cbot); \ctrl{7.45}{\cmid} \targ{7.45}{\cbot}

\gatebox{11.80}{\ctop}{2.35cm}{$R_Y(\theta_{S+1})$}
\gatebox{11.80}{\cmid}{2.35cm}{$R_Y(\theta_{S+2})$}
\gatebox{11.80}{\cbot}{2.35cm}{$R_Y(\theta_{2S})$}
\node at (7.12,-0.40) {(c)};

\end{tikzpicture}}
    \caption{Plots include (a) the schematic of quantum neural networks; (b) circuit details for initial state preparation, including Hadamard gates, CNOT-gates, and RZ-gates; and (c) circuit architecture for the variational model, including RY-gates and CNOT-gates.}
    \label{fig5}
\end{figure}

In classical computing, neural networks are widely employed for predictive learning tasks, including power system stability assessment \cite{an2020data}, state estimation \cite{tian2021neural}, and so on. 
In this context, the quantum counterpart of classical deep learning, with quantum neural networks (QNNs) as their basic components, has attracted significant research interest. 
The schematic of QNNs is shown in Fig. \ref{fig5}, comprising initial state preparation, a variational model, and measurements. 
Yu \emph{et al.} \cite{yu2023provable} demonstrate that parameterized quantum circuits, as a representation of QNNs, can perform functions analogous to classical counterparts. 
A comparative analysis of model complexity between parameterized quantum circuits and classical neural networks for approximating high-dimensional smooth functions reveals that parameterized quantum circuits require an exponential reduction in the number of parameters, provided that the target function meets specific smoothness conditions. 
This finding provides the first rigorous mathematical demonstration of the quantum advantage for parameterized quantum circuits in function approximation. 
By demonstrating this quantum advantage, this work advances quantum AI theory, paving the way for quantum solutions to be applied across various domains, including predictive tasks. 
As demonstrated in \cite{abbas2021power}, QNNs exhibit enhanced expressive power and higher parameter efficiency over their classical counterparts. 

Among the predictive tasks, image classification is a typical application. 
Its fundamental objective is to learn the mapping between input image features and class labels from labeled data, then predict the class labels of unseen images. 
Sebastianelli \emph{et al.} \cite{sebastianelli2021circuit} introduce a hybrid quantum-classical convolutional neural network architecture for remote sensing imagery classification, integrating QNNs into classical convolutional networks. 
In this model, the output of the preceding classical fully connected layer is encoded as rotation angles of quantum gates, thereby mapping the information into a high-dimensional Hilbert space. 
The output of the quantum layer, obtained by quantum measurements, is then passed to the subsequent classical layers for classification. 
Experimental results indicate that its classification accuracy surpasses that of a structurally similar classical model. 
However, as the number of QNN layers increases, it faces challenges such as high training complexity and declining accuracy. 
Therefore, Wang \emph{et al.} \cite{wang2024shallow} designed a shallow QNN. 
The core idea is to use Pauli-X rotation gates to modulate qubit amplitudes, Pauli-Z rotation gates to adjust qubit phases, and entanglement gates to establish correlations between qubits. 
This approach effectively limits both circuit depth and parameter count, thereby circumventing the gradient vanishing problem common to deeper quantum circuits. 
Notably, all the aforementioned models are, essentially, adaptations of classical neural networks. 
In addition, Gong \emph{et al.} \cite{gong2024quantum} propose a purely quantum architecture for image classification. 
The results demonstrate that this model converges faster than its classical counterpart. 

In addition, quantum predictive learning has also been applied to power systems. 
To address the challenges posed by high-dimensional, nonlinear features in transient stability assessment, a QNN-based method is proposed in \cite{zhou2022noise}. 
The algorithm incorporates a high expressivity, low-depth quantum circuit, enabling the effective extraction of transient stability characteristics. 
Additionally, quantum natural gradient descent is used to efficiently train the parameters of quantum circuits. 
The experimental results show that the algorithm can not only predict system stability, but also provide confidence intervals. 
Even for the IEEE 300-bus system, the quantum algorithm achieves an accuracy above $0.95$, whereas the classical algorithm remains below $0.95$. 
This highlights the superior performance of the quantum algorithm over its counterpart. 

In light of the above, existing studies report several potential benefits of quantum predictive learning, including improved prediction accuracy, faster convergence, and reduced parameter requirements. 
However, these results correspond to different models, datasets, tasks, and experimental settings and therefore do not constitute evidence of a universal QNN advantage. 
Meanwhile, several issues should be carefully considered. 
First, repeated circuit evaluations for gradient estimation and parameter optimization may introduce considerable training overhead. 
Increasing circuit depth may not only improve model capability, but also may increase computational cost and optimization difficulty, while deeper circuits are more susceptible to accumulated errors on current quantum hardware. 
In addition, quantum measurements generally require repeated circuit executions to obtain reliable expectation values, and the resulting sampling overhead can increase substantially with the number of parameters and required measurement precision. 
Gate errors, decoherence, and measurement errors caused by hardware noise may affect model outputs and training stability, particularly for deeper circuits. 
Moreover, reported QNN advantages are often obtained using different datasets, architectures, and training settings. 
Therefore, fair comparisons require consistent evaluation protocols and appropriately optimized classical baselines. 
Finally, as the problem size increases, QNNs may require more qubits and deeper circuits. Consequently, current studies remain largely limited to small-scale power systems. 

Meanwhile, the low dimensionality of the final discriminant output can reduce the resources required for measurement and readout. 
Here, it is worth noting that the overall quantum resource requirements are not solely determined by the output dimension, but also depend on factors such as input feature dimensionality, encoding strategy, quantum circuit width and depth, measurements, and parameter optimization. 
Furthermore, during training, the cost mainly arises from repeated executions of quantum circuits, gradient estimation, and parameter optimization. 
In contrast, the inference cost refers to the resources required for executing the trained quantum circuit, including quantum state evolution and measurement operations. 
For power system tasks such as stability assessment, the low-dimensional output mainly reduces the complexity of quantum measurement and classification result readout, while the costs associated with the number of qubits, circuit executions, and parameter optimization still depend on the input features, circuit architecture, and required prediction accuracy. 
Nevertheless, quantum predictive learning is expected to provide new perspectives on stability assessment, dynamic security assessment, anomaly classification, and fault diagnosis, among others.

Current studies on quantum predictive learning mainly rely on classical and noisy simulations, while some small QNNs have also been tested on real quantum hardware. 
In the near term, shallow QNNs can be integrated with classical neural networks and implemented on NISQ devices. 
Such models can process small datasets for stability assessment. 
In the medium term, increased qubit numbers and improved gate fidelity may support larger QNNs with more input features. 
These models may be extended to medium-scale power systems and more complex operating conditions. 
In the long term, fault-tolerant quantum computers may support deeper QNNs and larger datasets. 
This could extend quantum predictive learning to large-scale power system tasks. 

\subsection{Quantum Generative Learning}
In recent years, quantum generative learning has become a prominent research focus, with quantum generative adversarial networks representing a notable achievement in this field. 

Dallaire-Demers and Killoran \cite{dallaire2018quantum} and Lloyd and Weedbrook \cite{lloyd2018quantum} pioneered the theoretical analysis of its feasibility. 
Furthermore, experimental verifications in \cite{situ2020quantum} and \cite{zoufal2019quantum} demonstrate that the construction of the generator in classical generative adversarial networks based on QNNs can effectively learn the features of training data, thus enabling the generation of data with highly similar characteristics. 
To enhance training stability, Chakrabarti \emph{et al.} \cite{chakrabarti2019quantum} utilized the Wasserstein distance to calculate the loss function. 
In addition, leveraging enhanced quantum generative adversarial networks, Zhu \emph{et al.} \cite{zhu2022generative} generated joint probability distributions, while Tsang \emph{et al.} \cite{tsang2023hybrid} produced high-resolution images. 
Beyond quantum generative adversarial networks, based on the hypothesis that time series can be modeled by quantum processes, Horowitz \emph{et al.} \cite{horowitz2022quantum} proposed a generative quantum model based on Hamiltonian learning and demonstrated its ability to capture features of high-dimensional temporal data through experimental verification. 
In power systems, the quantum-classical hybrid generative adversarial network architecture is proposed for renewable scenario generation in \cite{11342337}, which consists of a quantum generator and a classical discriminator. 
Experimental results demonstrate that the quantum-enhanced model achieves better accuracy than its classical counterpart. 
Taking the wind power scenario as an example, the quantum model achieves a Wasserstein distance of $0.6889$, compared with $0.6896$ for transformer GANs and $0.7861$ for LSTM, demonstrating that the scenarios generated by the quantum algorithm are of higher quality. 
This advantage is also evident in small-sample learning \cite{11342337}. 

In sum, quantum generative learning has long attracted significant research interest, due not only to its ability to generate higher-quality samples but also to its superior model generalization performance under limited training samples, as illustrated in \cite{huang2025generative}. 
Leveraging these advantages, generative quantum models can be further applied to renewable scenario generation and time-series forecasting in power systems. 
However, an issue lies in the typically high dimensionality of the generated samples, which current quantum computing resources struggle to handle, thereby limiting scalability. 

Current evidence for quantum generative learning mainly comes from classical and noisy simulation. 
In the near term, hybrid quantum–classical generative models are more practical for NISQ devices. 
They can be applied to renewable scenario generation and data augmentation using small or compressed datasets. 
In the medium term, larger and more reliable quantum processors may support multivariate scenario generation. 
Such models may capture more complex temporal and spatial correlations in renewable generation and load data. 
In the long term, fault-tolerant quantum computers may support high-dimensional quantum generative models. 
These models could enable system-level scenario generation for large power systems and spatiotemporal datasets. 

\subsection{Quantum Probabilistic Inference}
\subsubsection{Quantum Bayesian}
Quantum Bayesian methods have made considerable progress in recent years. 
In 1997, Tucci \cite{tucci1997quantum} pioneered a quantum Bayesian network framework that extends classical Bayesian networks to the quantum realm by replacing classical probabilities with complex probability amplitudes. 
It models the conditional probabilities of quantum systems by propagating and interfering quantum amplitudes across nodes, enabling accurate modeling of quantum uncertainties. 

Subsequent research has demonstrated the significant computational advantages of this framework. 
It was noted in \cite{low2014quantum} that the amplitude amplification algorithm can accelerate the Bayesian inference process, while Berner \emph{et al.} \cite{berner2021quantum} further demonstrated that such quantum speedup is applicable to classical sampling of posterior distributions. 

This quantum Bayesian framework has the potential to open new avenues for power system uncertainty analysis, state estimation, parameter inference, and model calibrations, to cite a few, which are identification tasks with high computational complexity \cite{blei2017variational}. 

\subsubsection{Quantum Boltzmann Machines}
In addition to quantum Bayesian models, quantum Boltzmann machines (QBMs) have attracted increasing attention as a quantum-enhanced method for learning complex probability distributions. 
As shown in \cite{amin2018quantum} and \cite{zoufal2021variational}, QBMs extend classical Boltzmann machines by introducing quantum fluctuations through transverse-field Hamiltonians. 
This mechanism enables the representation of probability distributions with rich multimodal structures, strong correlations, and non-Gaussian characteristics. 
The advantage arises from the ability of quantum sampling processes to explore rugged energy landscapes more efficiently by exploiting superposition and tunneling effects, thus facilitating faster convergence toward equilibrium distributions. 

For power systems, which exhibit high-dimensional uncertainty arising from renewable resources, load variability, and stochastic disturbances, QBMs offer a promising approach to uncertainty-aware modeling and probabilistic inference. 
QBMs can approximate complex stochastic patterns more expressively than classical generative models \cite{coopmans2024sample}. 
Furthermore, hybrid training frameworks allow QBMs to be implemented on near-term quantum hardware by combining quantum Gibbs-state sampling with classical parameter optimization \cite{torlai2017neural}. 
This enables practical deployment even on noisy devices. 

Within the roadmap for the AI-enabled evolution of power system analytics, QBMs have the potential to play an important role for applications such as scenario generation, probabilistic security assessment, and risk-aware decision-making \cite{perdomo2018opportunities, liu2026quantum}. 
By improving the representation of complex high-dimensional uncertainties, QBMs provide a foundation for incorporating quantum-assisted probabilistic analysis into future power system applications. 

In the near term, small Bayesian models and hybrid QBMs can be explored on NISQ devices. 
They can support uncertainty analysis and parameter inference for simplified power system cases. 
In the medium term, larger quantum processors may enable more stochastic variables and more complex probability distributions to be considered. 
In the long term, fault-tolerant quantum computers may enable large-scale Bayesian inference and quantum sampling. 
These techniques could support probabilistic state estimation and system-wide risk assessment for large power systems. 

\subsection{Quantum Reasoning}
Recent studies have begun to explore the complementary roles of large language models (LLMs) and quantum computing. 
One representative direction is the use of LLMs to automate the formulation of quantum-compatible optimization models. 
For example, the LLM-QUBO framework demonstrates that natural-language problem descriptions can be systematically translated into QUBO formulations via LLM-driven reasoning and constraint parsing. 
This approach reduces the expertise barrier in quantum modeling and enables non-specialists to interact with quantum computing backends through high-level problem descriptions, while preserving the mathematical structure required by quantum solvers \cite{zhang2025llm}. 
However, the model reliability still requires rigorous verification. 
First, LLMs may omit critical constraints, such as power balance, ramping, reserve, and network security constraints, or introduce mathematical errors and dimensional inconsistencies, resulting in models that deviate from actual power systems. 
Second, QUBO formulations handle constraints through penalty terms. 
Insufficient penalty coefficients may lead to infeasible solutions, whereas excessively large coefficients may increase the coefficient dynamic range and make the problem more difficult to solve. 
In addition, LLMs remain susceptible to hallucinations, prompt sensitivity, and limited model traceability. 
Therefore, a more reliable workflow should verify constraint completeness, mathematical consistency, and physical feasibility before quantum solving, followed by power flow analysis and security-constraint validation of candidate solutions. 

A second emerging direction emphasizes quantum-assisted learning within classical AI workflows, particularly through knowledge distillation. 
Recent work on quantum relational knowledge distillation shows that quantum feature spaces and quantum kernel evaluations can be used during training to enhance the transfer of relational knowledge between LLMs \cite{liu2025quantum}. 
In this paradigm, quantum computation is introduced as a training-time enhancement rather than an inference-time requirement, allowing the distilled models to remain fully classical at deployment. 

In summary, LLMs improve accessibility and problem formulation for quantum computing, while quantum resources enrich representation and learning during critical training stages. 
This offers a viable pathway for intelligent decision support in complex power system scenarios. 
However, quantum reasoning is still in an exploratory stage. Current evidence mainly comes from theoretical studies and proof-of-concept experiments. 

In the near term, classical AI can assist in QUBO formulation, while small quantum processors can solve reduced optimization problems. 
This framework can be explored for small-scale scheduling and resource allocation tasks. 
In the medium term, improved quantum hardware may allow AI models to interact with larger quantum optimization and reasoning modules. 
This may support more complex operational decisions in medium-scale power systems. 
In the long term, fault-tolerant quantum computers may enable deeper integration with advanced AI models. 
Such systems could handle larger-scale uncertainties and operational constraints. 
They may support system-level scheduling, planning, and intelligent decision making. 

\subsection{Outlook}
In conclusion, by leveraging quantum physical properties, such as superposition and entanglement, quantum technologies hold the potential to alleviate the computational bottlenecks encountered by classical AI in power system tasks, as demonstrated in the following aspects. 

$(1)$ For geometric structure optimization of power system data, the large-scale integration of renewable energy sources and measurement devices results in massive and high-dimensional datasets. When classical AI methods are employed for dimensionality and data analysis, the associated computational cost can increase substantially. By exploiting quantum states to encode high-dimensional information, quantum computing offers the potential to alleviate the high computational complexity. 

$(2)$ For power system tasks that involve feature learning, such as fault diagnosis, load forecasting, and anomaly detection, models are required to extract informative features from raw measurement data to effectively characterize system operating states. 
The introduction of quantum computing maps classical data into the quantum space, providing new approaches for complex system feature representation. 
Quantum-enhanced methods may provide potential advantages for improving model convergence performance and result accuracy, where such improvements depend on specific model architectures and feature mapping strategies, requiring validation for specific power system tasks.

$(3)$ For power system tasks, such as stability assessment, training samples associated with rare events, including extreme faults and unstable conditions, are typically limited. 
Quantum-enhanced methods may provide potential benefits for improving model generalization under limited-sample conditions, which requires experimental validation for specific tasks.

The specific outlook is manifested in the following two aspects.
First, QNN training currently relies on the parameter-shift rule, which suffers from low gradient-estimation efficiency, resulting in substantial computational overhead. 
Consequently, a promising direction is to develop more efficient and direct training algorithms tailored to specific power system scenarios. 
Second, quantum AI demonstrates the potential for uncertainty modeling, scenario reduction, risk assessment, and real-time decision-making in power systems. 
This may provide more efficient support for security assessment and dispatch control under the high penetration of renewable energy. 
This calls for the design of quantum-classical hybrid architectures to enhance the scalability of quantum AI algorithms and meet practical needs. 

Now, we next proceed to summarize the current quantum hardware and simulators to gain a clearer understanding of the present stage of quantum computing development.

\section{Quantum Computing Hardware and Platforms for Applications}
This section reviews the main quantum hardware architectures and cloud platforms and examines the associated opportunities and limitations for their applications in power systems. 

\begin{table*}[htbp]
    \centering
    \caption{Analysis of Major Quantum Computing Hardware Technologies}
    \begin{tabular}{ccc}
    \toprule
    \textbf{Route} & \textbf{Application Strengths} & \textbf{Limitations}  \\   \hline
    Superconducting & \ding{172} Good programmability \ding{173} Suitable for QAOA and variational circuits
     & \ding{172} Short coherence time \ding{173} Scaling challenges \\  \hline
    Photonic & \makecell{\ding{172} High scalability \ding{173} Robust to noise \\ \ding{174} Suitable for combinatorial optimization}
     & \makecell{\ding{172} Limited programmability \\ \ding{173} Not ideal for circuit-based algorithms} \\  \hline
     Ion-trap & High fidelity and precision
     & \ding{172} Slow operation \ding{173} Limited scalability \\  \hline
     Neutral-atom & \ding{172} Large qubit numbers \ding{173} Suitable for Ising-type models  & Stability and control complexity \\ 
     \bottomrule     
    \end{tabular}
    \label{table3}
\end{table*}

\subsection{Major Technological Routes for Quantum Computing Hardware}
The rapid progress of quantum computing over the past decade has been driven primarily by advances in hardware design. 
Current development efforts are centered on four dominant technological routes \cite{de2021materials,NAP25196}: superconducting-, photonic-, ion-trap-, and neutral-atom-based quantum computing architectures. 
In addition to these dominant approaches, semiconductor-spin-based architectures, actively developed by Intel \cite{intel2023chip}, and topological-based architectures, advanced by Microsoft \cite{microsoft2024topological}, are also gaining momentum as promising alternatives. 
An overview is provided in Table \ref{table3}. 

Regarding the selection of hardware, large-scale optimization problems, such as unit commitment, can be addressed using photonic-based hardware. 
This is because the hardware architecture can support a large number of qubits, and these optimization problems can be formulated as Ising models. 
Power system tasks with stringent accuracy requirements, such as simulation, may be suited to ion-trap-based hardware due to their higher fidelity. 
In addition, quantum AI-related tasks can be implemented on superconducting-based hardware, which features relatively mature control technologies and is well suitable for variational quantum circuits.

\subsection{Cloud-Based Quantum Computing Platforms}
Due to the high construction cost, stringent operating conditions, and substantial maintenance requirements of quantum computers, local deployment remains uncommon. 
As a result, cloud-based access has become the primary method of utilizing NISQ hardware \cite{golec2024quantum}. 
Through cloud platforms, users can access quantum processors developed by various companies and research institutions, greatly reducing the barrier to experimenting with new quantum algorithms. 

At present, quantum cloud service providers are mainly concentrated in the United States, China, and Europe. 
Each region has formed its own technical ecosystem that reflects the hardware route pursued by local companies and research laboratories. 
Table \ref{table4} presents a regional comparison of the leading quantum cloud platforms, highlighting their providers, hardware types, and qubit scales \cite{ibmquantum, google2025quantum, rigetti_qcs, aws_braket, microsoft_azure_quantum, dwave_leap, xanadu_cloud, qutech_quantum_inspire, oqc_qcaas, pasqal_cloud, quandela_cloud, ctq_tianyan, quantumctek_cloud_en, baqis_quafu, arclight_cloud, origin_cloud, huayi_cloud, spinq_cloud, cmcc_wuyue_quantum, qboson_cloud}.

\begin{table*}[htbp]
    \centering
    \caption{Analysis of Major Quantum Computing Hardware Technologies}
    \begin{tabular}{cccccc}
    \toprule
    \textbf{Area} & \textbf{Platform} & \textbf{Country} & \textbf{Providers} & \textbf{Hardware} & \textbf{Max Number of Qubits}  \\   \hline
    \multirow{14}{*}{\makecell{North \\ America}} & IBM Q Experience
     & USA & IBM & superconducting-based & 433 \\  
    & \multirow{3}{*}{Google Cloud} & \multirow{3}{*}{USA} & Google, Rigetti & superconducting-based & 80 \\
    &  &  & AQT, IonQ & ion-trap-based & 20 \\
    &  &  & Pasqal & neutral-atom-based & 100 \\
    & Rigetti QCS & USA & Rigetti & superconducting-based & 84 \\
    & \multirow{4}{*}{Amazon Braket} & \multirow{4}{*}{USA} & OQC, Rigetti & superconducting-based & 80 \\
    &  &  & IonQ & ion-trap-based & 25 \\
    &  &  & Xanadu & photonic-based & 216 \\
    &  &  & QuEra & neutral-atom-based & 256 \\
    & \multirow{3}{*}{Microsoft Azure  Quantum} & \multirow{3}{*}{USA} & IonQ & ion-trap-based & 23 \\
    & & & Quantinuum, QCI, Rigetti & superconducting-based & 40 \\
    & & & Pasqal & neutral-atom-based & 100 \\
    & D-Wave Leap & Canada & D-Wave & quantum-analog-based & $5000+$ \\
    & Xanadu Cloud & Canada & Xanadu & photonic-based & 216 \\ \hline
    \multirow{6}{*}{Europe} & QuTech Quantum & \multirow{2}{*}{Netherland} & \multirow{2}{*}{Qutech} & silicon-spin-based & 2 \\
    & Inspire &  &  & superconducting-based & 5 \\
    & Computing-as-a-Service & U.K. & Oxford Quantum Circuits & superconducting-based & 4 \\
    & Pasqal Cloud & \multirow{2}{*}{France} & \multirow{2}{*}{Pasqal} & \multirow{2}{*}{neutral-atom-based} & \multirow{2}{*}{100} \\
    & Services &  &  &  &  \\
    & Quandela Cloud & France & Quandela & photonic-based & 6 \\ \hline
    \multirow{19}{*}{Asia} & China Telecom "Tianyan" & China & QuantumCTek & superconducting-based & 504 \\
    & QuantumCTek & China & \makecell{CAS Institute of Quantum \\ Information and Quantum \\ Technology Innovation} & superconducting-based & 294 \\
    & Quafu Quantum Cloud & China & \makecell{Beijing Academy of Quantum \\Information Sciences, Institute of \\ Physics CAS, Tsinghua University} & superconducting-based & 156 \\
    & Huguang Quantum Cloud & China & China Telecom Quantum & superconducting-based & 66 \\
    & Origin Quantum Cloud & China & Origin Quantum & superconducting-based & 180 \\ 
    & Huayi Quantum Cloud & China & Huayi Quantum & ion-trap-based & 11 \\ 
    & \multirow{2}{*}{SpinQ Cloud} & \multirow{2}{*}{China} & \multirow{2}{*}{SpinQ Technology} & nuclear-magnetic-resonance & 5 \\
    & & & & superconducting-based & 8 \\
    & \multirow{5}{*}{\makecell{China Mobile “Wuyue” \\ Quantum Cloud}} & \multirow{5}{*}{China} & \multirow{5}{*}{\makecell{Yangtze River Delta Industrial \\ Innovation Center, Institute of \\ Physics CAS, Beijing Academy of \\ Quantum Information Sciences, \\ Qike Quantum}} & \multirow{5}{*}{\makecell{superconducting-based \\ ion-trap-based}} & \multirow{5}{*}{\makecell{429(sum of four devices) \\ 20}} \\
    & & & &  &  \\ 
    & & & &  &  \\
    & & & &  &  \\
    & & & &  &  \\
    & Quantum Cloud & China & CAS Cold Atom & neutral-atom-based & 15 \\
    & Qboson Cloud & China & Qboson Quantum & Coherent Ising machine & 3000 \\       
     \bottomrule     
    \end{tabular}
    \label{table4}
\end{table*}

Despite the rapid development of global quantum cloud platforms, a noticeable gap remains between the capabilities of available quantum hardware and the computational requirements of practical power system applications. 
Most cloud-accessible devices still suffer from limited qubits, restricted connectivity, and non-negligible noise levels. 
Furthermore, variations in hardware architectures across platforms complicate the development of portable and hardware-agnostic quantum workflows. 
Even so, continued improvements in device fidelity, the emergence of larger processors, and the maturation of cloud-native quantum development environments provide a promising foundation for future progress. 
As quantum hardware evolves, cloud platforms are expected to facilitate the convergence of quantum computing and AI, paving the way for real-world power system applications. 

\subsection{Quantum Simulator}
Limited by current quantum hardware, quantum simulators have become an ideal platform for developing and validating quantum algorithms. 
In essence, quantum simulators leverage classical computers to emulate quantum processes, including superposition and entanglement. 
Several quantum simulators have been widely adopted, with popular examples including IBM’s Qiskit \cite{qiskit2024}, Pennylane \cite{bergholm2018pennylane}, and TensorCircuit \cite{zhang2023tensorcircuit}, to name a few. Detailed descriptions are provided below. 

Qiskit is one of the most widely used quantum development platforms, with its advantage lying in a comprehensive software ecosystem. 
It also has strong noise modeling capabilities, making it suitable for error analysis and validation. 

Pennylane is better suited to quantum AI algorithms. 
It supports automatic differentiation, which enables efficient training of variational quantum algorithms. 
Moreover, Pennylane facilitates the construction of hybrid quantum-classical models, which is a current mainstream direction. 
Nevertheless, the simulator focuses more on algorithmic aspects, while offering less control over the underlying hardware. 

TensorCircuit is a high-performance simulator that has emerged in recent years, with tensor network acceleration as its core concept. 
It is highly integrated with deep learning frameworks and well-suited to large-scale variational circuits. 
However, its ecosystem has not yet matured. 

It is worth noting that quantum simulators fundamentally rely on classical computing resources. 
Therefore, as the number of qubits increases, the demand for classical computing resources becomes unimaginable. 
In sum, as essential research tools, quantum simulators are advancing quantum computing technology. 
They will inevitably play a crucial role in the practicalization of quantum computing.

\section{Challenges and Future Research}
In this section, we discuss current challenges in the application of quantum algorithms to power systems and outline future research directions. 

\subsection{Current Challenges}
Despite rapid progress in quantum algorithms and hardware, quantum computing for power systems remains in its early stages, with substantial gaps between theoretical potential and practical applicability. 
Unlike general computational tasks, power system applications are governed by specific physical characteristics, such as large-scale interconnected networks, and real-time operational requirements. 
These characteristics amplify the limitations of current quantum computing and should be explicitly considered. 

\subsubsection{Physical Bottleneck}
Current quantum devices remain in the NISQ era, characterized by limited connectivity, short coherence times, and high noise levels \cite{chen2023complexity}. 
These physical constraints prevent the realization of fault-tolerant quantum computation and severely limit the depth and complexity of executable quantum circuits. 
This limitation is particularly critical for power systems, where physical quantities such as power flows should be calculated with sufficient accuracy to ensure reliable system operation. 
Errors caused by quantum noise may propagate through the interconnected network and affect the accuracy of subsequent system analysis. 
As a result, the stringent accuracy and reliability requirements imposed by the physical operation of power systems amplify the impact of current quantum hardware limitations. 

\subsubsection{Scalability Bottleneck}
In the current quantum hardware landscape, the limited availability of qubits has emerged as a major barrier to the practical application of quantum algorithms. 
This issue becomes particularly significant for power systems because power grids consist of a large number of physically interconnected buses, generators, loads, transmission lines, and measurement devices, among others. 
As the physical scale of the network increases, the number of variables and constraints required to represent its operating state increases accordingly. 
The resulting problem dimension may rapidly exceed the number of qubits available on current quantum devices. 
Therefore, the continuously expanding physical scale and complexity of power networks impose stringent scalability requirements on quantum computing. 

\subsubsection{Theoretical-Optimality Bottleneck}
Quantum linear solvers with provable asymptotic advantages, such as the HHL algorithm, remain largely impractical on current hardware. 
We would like to emphasize here that the advantages of the HHL algorithm still remain at the theoretical level, as they rely on ideal assumptions. 
This issue is particularly relevant to power systems, whose operating states are determined by a large number of continuously varying physical quantities, including voltage magnitudes, and active power. 
First, regarding the input bottleneck, encoding these classical physical quantities into quantum states incurs a high computational cost. 
Second, concerning the output problem, the HHL algorithm outputs a quantum state rather than direct numerical values. 
Consequently, repeated measurements on quantum states are required. 
Third, the noise inherent in quantum devices undermines numerical precision. 
Therefore, although quantum algorithms may offer theoretical advantages, these advantages cannot yet be fully validated under the physical and operational requirements of practical power systems. 

\subsubsection{Encoding Bottleneck}
Encoding power system topology or data into quantum states requires a certain cost, with complexity scaling with problem size. 
This challenge is particularly relevant because power systems simultaneously contain numerous physical information, including network topology, voltage measurements, and temporal operating data. 
Among current encoding methods, amplitude encoding requires fewer qubits for representing high-dimensional power system data but is relatively difficult to implement, whereas angle encoding is easier to implement but requires greater quantum resources as the dimensionality of system data increases. 
Consequently, efficiently representing large-scale physical information constitutes a major encoding bottleneck for quantum computing in power systems. 

\subsubsection{Quantum Parameter Optimization Bottleneck}
On the one hand, some quantum circuits are prone to barren plateaus during training \cite{larocca2025barren}, thereby impeding parameter optimization. 
On the other hand, the parameter gradients required for training are typically computed using the parameter-shift rule, which requires the construction and execution of two distinct circuits per parameter, severely limiting training efficiency. 
This issue is important for power systems because many simulation and monitoring tasks require rapid computation. 
Take power system simulation as an example: even NISQ-compatible approaches, such as VQLS, encounter the barren plateau problem when applied to transient or static simulation tasks. 
In addition, the computational burden of gradient calculation limits the simulation's real-time capability. 

\subsubsection{Variational Quantum Algorithm Bottleneck}
We would like to state that variational methods are not guaranteed to possess a quantum advantage; rather, they are proposed as a practically realizable approach for the NISQ era. 
Essentially, variational methods are quantum-classical hybrid optimization frameworks in which the quantum part constructs the target state, while the classical part optimizes the parameters. 
Compared with general quantum algorithms, variational algorithms are more robust to noise and impose less demanding hardware requirements, thereby lowering the barrier to practical implementation. 
In this context, various forms have been developed, such as VQLS, QAOA, and QNNs, as introduced previously. 
Fortunately, owing to the unique quantum physical properties, variational quantum circuits exhibit superior expressive power, which has been demonstrated in \cite{du2020expressive}. 
However, the unique advantages of variational quantum algorithms have yet to be fully demonstrated in power system applications. 

\subsubsection{Power System Structure Bottleneck}
In addition to the above challenges, the physical structure of power systems imposes specific constraints on quantum algorithms. 
First, power networks are generally sparse because each bus is physically connected to a limited number of neighboring buses. 
Consequently, network-related matrices, such as admittance matrices, are sparse. 
This property is potentially favorable for quantum linear solvers such as HHL, whose theoretical complexity depends on the matrix sparsity. 
However, exploiting this advantage requires efficient quantum state preparation and measurement readout. 
Otherwise, the associated overhead may offset the theoretical speedup. 
Second, the admittance matrix exhibits structural symmetry. 
Nevertheless, such matrices may not directly satisfy the mathematical requirements of specific quantum linear solvers. 
In this context, additional transformations are necessary, which introduce additional quantum resources. 

\subsection{Future Research}
In response to bottlenecks, future research can focus on the following directions. 

\subsubsection{Noise Mitigation}
To enhance the reliability and robustness of quantum measurement results under noisy hardware conditions, developing noise-resilient quantum algorithms, error-suppression strategies, and improved sampling procedures will be essential for security-constrained and resilience-oriented problems. 
Specifically, a calibration method needs to be applied to the recorded quantum error data points to estimate and remove the systematic bias. 
Furthermore, advanced filtering methods, such as the discrete wavelet transform, should be applied to mitigate multifractal errors while preserving the true values \cite{saghafi2025multifractality}. 

\subsubsection{Physical-Informed Quantum Design}
To simplify quantum architectures, exploiting the intrinsic physical characteristics of power systems is meaningful. 
For example, by leveraging the admittance matrix properties, such as symmetry \cite{glover2002power}, the corresponding circuit depth, gate count, and the number of parallel executions required can be significantly reduced, thereby enabling the design of quantum solvers that meet real-time requirements. 
In other power system problems, embedding physical information into quantum circuits can enhance the accuracy of the results. 
In addition, as noted in \cite{siegl2024solving}, embedding physical priors directly into the variational ansatz can also help alleviate barren plateaus, thus improving training stability and efficiency on NISQ hardware. 
To achieve the above, it is necessary to explore the physical characteristics of power systems and appropriately devise the embedding rules for such information. 

\subsubsection{Efficient Encoding Strategy}
To achieve efficient encoding, it is essential to investigate dedicated encoding schemes for power systems to significantly reduce the complexity of mapping high-dimensional data to quantum states. 
For instance, amplitude encoding can be employed to encode high-dimensional data into quantum states through global normalization. 
In contrast, angle encoding can be applied to a selected set of local variables by parameterizing the rotations of quantum gates. 
Alternatively, these two encoding mechanisms can be jointly incorporated to address more complex data structures. 

\subsubsection{Quantum Circuit Optimization}
To mitigate the barren plateau issue, designing shallow quantum circuits with a modular architecture that retains the expressive capacity needed to capture the complex characteristics of power systems is worthwhile. 
In addition, employing strategies such as parameter sharing and local parameterization shows promise in substantially reducing circuit depth without compromising model performance. 
Meanwhile, to address the computational overhead for gradient estimation, improved approximate gradient methods or randomization-based parameter update strategies can be explored. 
These approaches would reduce the need for repetitive circuit executions, thereby enhancing training efficiency. 

\subsubsection{Scalability Enhancement}
To address the scalability challenges of quantum algorithms, adopting a hybrid framework that integrates classical neural networks with quantum circuits is promising. 
Specifically, classical neural networks are used to extract features and reduce dimensionality in massive power system datasets. 
The resulting lower-dimensional data are then processed by quantum circuits, effectively alleviating quantum resource constraints. 

\subsubsection{Quantum-Classical Hybridization}
To optimize computing resources, adopting a quantum-classical hybridization strategy is essential. 
Its core idea lies in the collaborative allocation of complex computational tasks. 
Specifically, quantum computers handle computationally intensive subtasks, while classical computers handle the remaining operations. 
Here, it is worth emphasizing that classical computing plays a supporting role within the strategy, enabling the adaptation of quantum algorithms to more complex application scenarios.
The strategy is, in fact, a transitional solution necessitated by current limited quantum computing resources. 
Taking the image classification task in \cite{wang2025hybrid} as an example, to reduce the depth of quantum circuits, a multi-channel parallel extraction module is constructed. 
In addition to the quantum channel, classical convolutional layers are introduced to extract image features. 
This design fully leverages the high-dimensional expressive power of quantum circuits and the large-scale processing capability of classical computing. 
Such structural flexibility enables handling more complex datasets, which is infeasible with purely quantum systems. 
Therefore, this collaborative computing paradigm can effectively alleviate the current shortage of quantum computing resources. 

Last but not least, current research of quantum computing in power systems is still at an early stage. 
Existing studies differ in terms of power system test cases, quantum hardware and simulators. 
A unified standardized benchmarking framework has yet to be established. 
Therefore, developing benchmarks based on standardized test cases will be a direction for future research. 

\section{Conclusion}
Quantum computing has emerged as a promising computational paradigm for addressing the increasing complexity of modern power systems. 
This survey reviewed the foundational principles of quantum computing, major algorithmic frameworks, and the applications in power system simulation, optimization, and AI-enabled analysis. 
Existing studies show that quantum techniques provide complementary capabilities in all three domains: VQA offers new possibilities for accelerating the solution of large-scale power system simulation models, quantum optimization approaches introduce alternative formulations for complex decision problems in dispatch and network operation, and AI-enabled quantum learning approaches enable novel mechanisms for extracting structure from high-dimensional data. 
Together, these advances illustrate the potential of quantum computing in supporting a future power system. 
At the same time, current applications remain restricted by hardware limitations, noise sensitivity, scalability challenges, and the absence of standardized benchmarks. 
Together, these insights outline both the progress achieved and the technical gaps that must be bridged. 

Looking ahead, integrating quantum technologies with AI-driven methodologies is expected to empower future power systems. 
Progress in quantum hardware and cloud platforms will enable quantum AI workflows that combine quantum hardware with advanced AI models for forecasting, control, and decision support. 
To realize this vision, future research must focus on developing scalable hybrid architectures, improving noise-robust quantum learning algorithms, and designing power system models that align with emerging quantum AI capabilities. 
These will provide new computational pathways for the operation and control of next-generation power systems.

\vspace{-0.32cm}
\newcommand{\BIBdecl}{\setlength{\itemsep}{0.02 em}}
\bibliographystyle{IEEEtran}
\bibliography{IEEEabrv,References.bib}

@IEEEtranBSTCTL{IEEEexample:BSTcontrol,
CTLuse_forced_etal       = "yes",
CTLmax_names_forced_etal = "5",
CTLnames_show_etal       = "1",
CTLdash_repeated_names = "no"
}

@ARTICLE{10799930,
  author={Chen, Zhe and Li, Zhihao and Lin, Da and Xie, Changjun and Wang, Zhewei},
  journal={Journal of Modern Power Systems and Clean Energy}, 
  title={Multi-Time-Scale Optimal Scheduling of Integrated Energy System with Electric-Thermal-Hydrogen Hybrid Energy Storage Under Wind and Solar Uncertainties},
  month=may,
  year={2025},
  volume={13},
  number={3},
  pages={904-914},
  doi={10.35833/MPCE.2024.000606}}

@ARTICLE{9275593,
  author={Cao, Di and Hu, Weihao and Zhao, Junbo and Zhang, Guozhou and Zhang, Bin and Liu, Zhou and Chen, Zhe and Blaabjerg, Frede},
  journal={Journal of Modern Power Systems and Clean Energy}, 
  title={Reinforcement Learning and Its Applications in Modern Power and Energy Systems: A Review},
  month=nov,
  year={2020},
  volume={8},
  number={6},
  pages={1029-1042},
  doi={10.35833/MPCE.2020.000552}}

@ARTICLE{10419164,
  author={Du, Pengbo and Huang, Bonan and Liu, Ziming and Yang, Chao and Sun, Qiuye},
  journal={Journal of Modern Power Systems and Clean Energy}, 
  title={Real-Time Energy Management for Net-Zero Power Systems Based on Shared Energy Storage},
  month=mar,
  year={2024},
  volume={12},
  number={2},
  pages={371-380},
  doi={10.35833/MPCE.2023.000535}}

@article{beg2018multi,
  title={Multi-physical electro-magnetic propulsion fluid dynamics: mathematical modelling and computation},
  author={B{\'e}g, O Anwar},
  journal={Mathematical Modelling: Methods, Applications and Research},
  pages={1--88},
  year={2018},
  publisher={Nova Science New York, USA}
}

@article{barros2025globally,
  title={Globally optimal sequencing of optimal reactive dispatch control adjustments to minimize operational losses in transmission systems by graph shortest path, parallel computing, and dynamic programming},
  author={Barros, Rafael Martins and Lage, Guilherme Guimar{\~a}es and Rab{\^e}lo, Ricardo de Andrade Lira},
  journal={European Journal of Operational Research},
  volume={320},
  number={1},
  pages={239--255},
  year={2025},
  publisher={Elsevier}
}

@article{pfaendler2024advancements,
  title={Advancements in quantum Computing—Viewpoint: Building adoption and competency in industry},
  author={Pfaendler, Sieglinde M-L and Konson, Konstantin and Greinert, Franziska},
  journal={Datenbank-spektrum},
  volume={24},
  number={1},
  pages={5--20},
  year={2024},
  publisher={Springer}
}

@article{cai2013experimental,
  title={Experimental quantum computing to solve systems of linear equations},
  author={Cai, X-D and Weedbrook, Christian and Su, Z-E and Chen, M-C and Gu, Mile and Zhu, M-J and Li, Li and Liu, Nai-Le and Lu, Chao-Yang and Pan, Jian-Wei},
  journal={Physical Review Letters},
  volume={110},
  number={23},
  note={{A}rt. no. 230501},
  year={2013},
  publisher={APS}
}

@book{nielsen2010quantum,
  title={Quantum Computation and Quantum Information},
  author={Nielsen, Michael A and Chuang, Isaac L},
  year={2010},
  publisher={Cambridge University Press}
}

@article{humble2019quantum,
  title={Quantum computing circuits and devices},
  author={Humble, Travis S and Thapliyal, Himanshu and Munoz-Coreas, Edgard and Mohiyaddin, Fahd A and Bennink, Ryan S},
  journal={IEEE Design \& Test},
  volume={36},
  number={3},
  pages={69--94},
  month=jun,
  year={2019},
  publisher={IEEE}
}

@book{braginsky1995quantum,
  title={Quantum Measurement},
  author={Braginsky, Vladimir B and Khalili, Farid Ya},
  year={1995},
  publisher={Cambridge University Press}
}

@article{roushan2020quantum,
  title={Quantum supremacy: Computational complexity and applications},
  author={Roushan, Pedram},
  journal={Bulletin of the American Physical Society},
  volume={65},
  year={2020},
  publisher={APS}
}

@article{zhou2022quantum,
  title={Quantum computing in power systems},
  author={Zhou, Yifan and Tang, Zefan and Nikmehr, Nima and Babahajiani, Pouya and Feng, Fei and Wei, Tzu-Chieh and Zheng, Honghao and Zhang, Peng},
  journal={IEnergy},
  volume={1},
  number={2},
  pages={170--187},
  month=jun,
  year={2022},
  publisher={TUP}
}

@article{harrow2009quantum,
  title={Quantum algorithm for linear systems of equations},
  author={Harrow, Aram W and Hassidim, Avinatan and Lloyd, Seth},
  journal={Physical Review Letters},
  volume={103},
  number={15},
  note={{A}rt. no. 150502},
  year={2009},
  publisher={APS}
}

@article{zaman2023step,
  title={A step-by-step {HHL} algorithm walkthrough to enhance understanding of critical quantum computing concepts},
  author={Zaman, Anika and Morrell, Hector Jose and Wong, Hiu Yung},
  journal={IEEE Access},
  volume={11},
  pages={77117--77131},
  year={2023},
  publisher={IEEE}
}

@article{cerezo2021variational,
  title={Variational quantum algorithms},
  author={Cerezo, Marco and Arrasmith, Andrew and Babbush, Ryan and Benjamin, Simon C and Endo, Suguru and Fujii, Keisuke and McClean, Jarrod R and Mitarai, Kosuke and Yuan, Xiao and Cincio, Lukasz and others},
  journal={Nature Reviews Physics},
  volume={3},
  number={9},
  pages={625--644},
  year={2021},
  publisher={Nature Publishing Group UK London}
}

@article{qi2024variational,
  title={Variational quantum algorithms: Fundamental concepts, applications and challenges},
  author={Qi, Han and Xiao, Sihui and Liu, Zhuo and Gong, Changqing and Gani, Abdullah},
  journal={Quantum Information Processing},
  volume={23},
  number={6},
  note={{A}rt. no. 224},
  year={2024},
  publisher={Springer}
}

@article{kandala2017hardware,
  title={Hardware-efficient variational quantum eigensolver for small molecules and quantum magnets},
  author={Kandala, Abhinav and Mezzacapo, Antonio and Temme, Kristan and Takita, Maika and Brink, Markus and Chow, Jerry M and Gambetta, Jay M},
  journal={Nature},
  volume={549},
  number={7671},
  pages={242--246},
  year={2017},
  publisher={Nature Publishing Group}
}

@article{bravo2023variational,
  title={Variational quantum linear solver},
  author={Bravo-Prieto, Carlos and LaRose, Ryan and Cerezo, Marco and Subasi, Yigit and Cincio, Lukasz and Coles, Patrick J},
  journal={Quantum},
  volume={7},
  note={{A}rt. no. 1188},
  year={2023},
  publisher={Verein zur F{\"o}rderung des Open Access Publizierens in den Quantenwissenschaften}
}

@article{golestan2023quantum,
  title={Quantum computation in power systems: An overview of recent advances},
  author={Golestan, Saeed and Habibi, MR and Mousavi, SY Mousazadeh and Guerrero, Josep M and Vasquez, Juan C},
  journal={Energy Reports},
  volume={9},
  pages={584--596},
  month=dec,
  year={2023},
  publisher={Elsevier}
}

@article{glover2022quantum,
  title={Quantum bridge analytics {I}: A tutorial on formulating and using {QUBO} models},
  author={Glover, Fred and Kochenberger, Gary and Hennig, Rick and Du, Yu},
  journal={Annals of Operations Research},
  volume={314},
  number={1},
  pages={141--183},
  year={2022},
  publisher={Springer}
}

@article{brush1967history,
  title={History of the {L}enz-{I}sing model},
  author={Brush, Stephen G},
  journal={Reviews of Modern Physics},
  volume={39},
  number={4},
  note={{A}rt. no. 883},
  month=oct,
  year={1967},
  publisher={APS}
}

@article{lucas2014ising,
  title={Ising formulations of many {NP} problems},
  author={Lucas, Andrew},
  journal={Frontiers in Physics},
  volume={2},
  note={{A}rt. no. 74887},
  year={2014},
  publisher={Frontiers}
}

@article{zhou2020quantum,
  title={Quantum approximate optimization algorithm: Performance, mechanism, and implementation on near-term devices},
  author={Zhou, Leo and Wang, Sheng-Tao and Choi, Soonwon and Pichler, Hannes and Lukin, Mikhail D},
  journal={Physical Review X},
  volume={10},
  number={2},
  pages={021067},
  year={2020},
  publisher={APS}
}

@article{king2025beyond,
  title={Beyond-classical computation in quantum simulation},
  author={King, Andrew D and Nocera, Alberto and Rams, Marek M and Dziarmaga, Jacek and Wiersema, Roeland and Bernoudy, William and Raymond, Jack and Kaushal, Nitin and Heinsdorf, Niclas and Harris, Richard and others},
  journal={Science},
  volume={388},
  number={6743},
  pages={199--204},
  year={2025},
  publisher={American Association for the Advancement of Science}
}

@article{wang2013coherent,
  title={Coherent Ising machine based on degenerate optical parametric oscillators},
  author={Wang, Zhe and Marandi, Alireza and Wen, Kai and Byer, Robert L and Yamamoto, Yoshihisa},
  journal={Physical Review A},
  volume={88},
  number={6},
  pages={063853},
  year={2013},
  publisher={APS}
}

@article{morstyn2024opportunities,
  title={Opportunities for quantum computing within net-zero power system optimization},
  author={Morstyn, Thomas and Wang, Xiangyue},
  journal={Joule},
  volume={8},
  number={6},
  pages={1619--1640},
  year={2024},
  publisher={Elsevier}
}

@article{cortes2018characterizing,
  title={Characterizing tolerable disturbances for transient-state safety in power networks},
  author={Zhang, Yifu and Cortés, Jorge},
  journal={IEEE Transactions on Network Science and Engineering},
  volume={6},
  number={3},
  pages={210--224},
  month={Jul.--Sep.},
  year={2018},
  publisher={IEEE}
}

@article{chen2024distributed,
  title={Distributed and distributionally robust chance-constrained scheduling for {VSC-MTDC} meshed {AC/DC} power networks},
  author={Chen, Xiao and Zhai, Junyi and Shen, Jingwen and Wang, Qingwei and Li, Shanying and Wang, Sheng},
  journal={IEEE Transactions on Network Science and Engineering},
  volume={11},
  number={4},
  pages={3317--3331},
  month={Jul.--Aug.},
  year={2024},
  publisher={IEEE}
}

@article{ajagekar2022quantum,
  title={Quantum computing and quantum artificial intelligence for renewable and sustainable energy: A emerging prospect towards climate neutrality},
  author={Ajagekar, Akshay and You, Fengqi},
  journal={Renewable and Sustainable Energy Reviews},
  volume={165},
  note={{A}rt. no. 112493},
  month=sep,
  year={2022},
  publisher={Elsevier}
}

@article{ajagekar2019quantum,
  title={Quantum computing for energy systems optimization: Challenges and opportunities},
  author={Ajagekar, Akshay and You, Fengqi},
  journal={Energy},
  volume={179},
  pages={76--89},
  month=jul,
  year={2019},
  publisher={Elsevier}
}

@article{feng2022distributed,
  title={Distributed networked microgrids power flow},
  author={Feng, Fei and Zhang, Peng and Zhou, Yifan and Wang, Lizhi},
  journal={IEEE Transactions on Power Systems},
  volume={38},
  number={2},
  pages={1405--1419},
  month=mar,
  year={2023},
  publisher={IEEE}
}

@article{zhou2020reachable,
  title={Reachable power flow},
  author={Zhou, Yifan and Zhang, Peng},
  journal={IEEE Transactions on Power Systems},
  volume={35},
  number={4},
  pages={3290--3293},
  month=jul,
  year={2020},
  publisher={IEEE}
}

@article{feng2020enhanced,
  title={Enhanced microgrid power flow incorporating hierarchical control},
  author={Feng, Fei and Zhang, Peng},
  journal={IEEE Transactions on Power Systems},
  volume={35},
  number={3},
  pages={2463--2466},
  month=may,
  year={2020},
  publisher={IEEE}
}

@ARTICLE{9110755,
  author={Feng, Fei and Zhang, Peng},
  journal={IEEE Transactions on Power Systems}, 
  title={Implicit $\text{Z}_\text{bus}$ {G}auss Algorithm Revisited},
  month=sep,
  year={2020},
  volume={35},
  number={5},
  pages={4108-4111},
  doi={10.1109/TPWRS.2020.3000658}}

@article{stott2007fast,
  title={Fast decoupled load flow},
  author={Stott, Brian and Alsac, Ongun},
  journal={IEEE Transactions on Power Apparatus and Systems},
  volume={PAS-93},
  number={3},
  pages={859--869},
  month=may,
  year={1974},
  publisher={IEEE}
}

@article{eskandarpour2020quantum,
  title={Quantum computing solution of DC power flow},
  author={Eskandarpour, Rozhin and Ghosh, Kumar and Khodaei, Amin and Zhang, Liuxi and Paaso, Aleksi and Bahramirad, Shay},
  journal={arXiv preprint arXiv:2010.02442},
  year={2020}
}

@article{feng2021quantum,
  title={Quantum power flow},
  author={Feng, Fei and Zhou, Yifan and Zhang, Peng},
  journal={IEEE Transactions on Power Systems},
  volume={36},
  number={4},
  pages={3810--3812},
  year={2021},
  publisher={IEEE}
}

@article{saevarsson2022quantum,
  title={Quantum computing for power flow algorithms: Testing on real quantum computers},
  author={S{\ae}varsson, Brynjar and Chatzivasileiadis, Spyros and J{\'o}hannsson, Hj{\"o}rtur and {\O}stergaard, Jacob},
  journal={arXiv preprint arXiv:2204.14028},
  year={2022}
}

@article{feng2023noise,
  title={Noise-resilient quantum power flow},
  author={Feng, Fei and Zhou, Yi-Fan and Zhang, Peng},
  journal={iEnergy},
  volume={2},
  number={1},
  pages={63--70},
  year={2023},
  publisher={TUP}
}

@article{gao2023solving,
  title={Solving {DC} power flow problems using quantum and hybrid algorithms},
  author={Gao, Fang and Wu, Guojian and Guo, Suhang and Dai, Wei and Shuang, Feng},
  journal={Applied Soft Computing},
  volume={137},
  note={{A}rt. no. 110147},
  month=apr,
  year={2023},
  publisher={Elsevier}
}

@article{joint2021odessa,
  title={Odessa disturbance},
  author={Joint, NERC and Texas, RE},
  journal={Odessa Disturbance},
  year={2021}
}

@book{watson2003power,
  title={Power Systems Electromagnetic Transients Simulation},
  author={Watson, Neville and Arrillaga, Jos and Arrillaga, Jos},
  volume={39},
  year={2003},
  publisher={Iet}
}

@article{zhang2010shifted,
  title={Shifted-frequency analysis for {EMTP} simulation of power-system dynamics},
  author={Zhang, Peng and Mart{\'\i}, Jos{\'e} R and Dommel, Hermann W},
  journal={IEEE Transactions on Circuits and Systems I: Regular Papers},
  volume={57},
  number={9},
  pages={2564--2574},
  month=sep,
  year={2010},
  publisher={IEEE}
}

@article{zhang2017key,
  title={Key techniques in real time digital simulation for closed-loop testing of {HVDC} systems},
  author={Zhang, Yi and Ding, Hui and Kuffel, Rick},
  journal={CSEE Journal of Power and Energy Systems},
  volume={3},
  number={2},
  pages={125--130},
  month=jun,
  year={2017},
  publisher={CSEE}
}

@book{hidary2021quantum,
  title={Quantum Computing: An Applied Approach},
  author={Hidary, Jack D},
  volume={1},
  year={2021},
  address = {Cham, Switzerland},
  publisher={Springer},
  edition = {2}
}

@article{zhou2021quantum,
  title={Quantum Electromagnetic Transients Program},
  author={Zhou, Yifan and Feng, Fei and Zhang, Peng},
  journal={IEEE Transactions on Power Systems},
  volume={36},
  number={4},
  pages={3813--3816},
  month=jul,
  year={2021},
  publisher={IEEE}
}

@article{duan2020survey,
  title={A survey on {HHL} algorithm: From theory to application in quantum machine learning},
  author={Duan, Bojia and Yuan, Jiabin and Yu, Chao-Hua and Huang, Jianbang and Hsieh, Chang-Yu},
  journal={Physics Letters A},
  volume={384},
  number={24},
  note={{A}rt. no. 126595},
  month=aug,
  year={2020},
  publisher={Elsevier}
}

@article{preskill2018quantum,
  title={Quantum computing in the {NISQ} era and beyond},
  author={Preskill, John},
  journal={Quantum},
  volume={2},
  pages={79},
  year={2018},
  publisher={Verein zur F{\"o}rderung des Open Access Publizierens in den Quantenwissenschaften}
}

@article{zhou2022noisy,
  title={Noisy-intermediate-scale quantum electromagnetic transients program},
  author={Zhou, Yifan and Zhang, Peng and Feng, Fei},
  journal={IEEE Transactions on Power Systems},
  volume={38},
  number={2},
  pages={1558--1571},
  month=mar,
  year={2023},
  publisher={IEEE}
}

@article{lou2026matrix,
  title={Matrix low-dimensional qubit casting based quantum electromagnetic transient network simulation program},
  author={Lou, Qi and Xu, Yijun and Gu, Wei},
  journal={IEEE Transactions on Quantum Engineering},
  year={2026},
  publisher={IEEE}
}

@article{liu2026quantum,
  title={A Quantum-Assisted Boltzmann Machine for Wind Power Probabilistic Learning},
  author={Liu, Chengjun and Xu, Yijun and Wen, Kai and Gu, Wei and Lu, Shuai and Mili, Lamine},
  journal={IEEE Transactions on Sustainable Energy},
  year={2026},
  publisher={IEEE}
}

@article{liu2026alleviating,
  title={Alleviating CoD in renewable energy profile clustering using an optical quantum computer},
  author={Liu, Chengjun and Xu, Yijun and Gu, Wei and Sun, Bo and Wen, Kai and Lu, Shuai and Mili, Lamine},
  journal={CSEE Journal of Power and Energy Systems},
  year={2026},
  publisher={CSEE}
}

@article{paterakis2023hybrid,
  title={Hybrid quantum-classical multi-cut benders approach with a power system application},
  author={Paterakis, Nikolaos G},
  journal={Computers \& Chemical Engineering},
  volume={172},
  note={{A}rt. no. 108161},
  month=apr,
  year={2023},
  publisher={Elsevier}
}

@article{ling2025hybrid,
  title={Hybrid quantum annealing decomposition framework for unit commitment},
  author={Ling, Jiajie and Zhang, Quan and Geng, Guangchao and Jiang, Quanyuan},
  journal={Electric Power Systems Research},
  volume={238},
  note={{A}rt. no. 111121},
  month=jan,
  year={2025},
  publisher={Elsevier}
}

@article{feng2022novel,
  title={Novel resolution of unit commitment problems through quantum surrogate {L}agrangian relaxation},
  author={Feng, Fei and Zhang, Peng and Bragin, Mikhail A and Zhou, Yifan},
  journal={IEEE Transactions on Power Systems},
  volume={38},
  number={3},
  pages={2460--2471},
  month=may,
  year={2023},
  publisher={IEEE}
}

@article{nikmehr2022quantum,
  title={Quantum distributed unit commitment: An application in microgrids},
  author={Nikmehr, Nima and Zhang, Peng and Bragin, Mikhail A},
  journal={IEEE transactions on power systems},
  volume={37},
  number={5},
  pages={3592--3603},
  year={2022},
  publisher={IEEE}
}

@article{yang2025scalable,
  title={A scalable fully distributed quantum alternating direction method of multipliers for unit commitment problems},
  author={Yang, Mingyu and Gao, Fang and Dai, Wei and Huang, Dejian and Gao, Qing and Shuang, Feng},
  journal={Advanced Quantum Technologies},
  volume={8},
  number={1},
  note={{A}rt. no. 2400286},
  year={2025},
  publisher={Wiley Online Library}
}

@article{fu2025quantum,
  title={Quantum-embedded robust optimization for resilience-constrained unit commitment},
  author={Fu, Wei and Xie, Haipeng and Chen, Chen and Bie, Zhaohong},
  journal={IEEE Transactions on Power Systems},
  volume={40},
  number={5},
  pages={3778--3792},
  month=sep,
  year={2025},
  publisher={IEEE}
}

@article{zheng2024fast,
  title={A fast quantum algorithm for searching the quasi-optimal solutions of unit commitment},
  author={Zheng, Xiaodong and Wang, Jianhui and Yue, Meng},
  journal={IEEE Transactions on Power Systems},
  volume={39},
  number={2},
  pages={4755--4758},
  month=mar,
  year={2024},
  publisher={IEEE}
}

@article{kaseb2024power,
  title={Power flow analysis using quantum and digital annealers: a discrete combinatorial optimization approach},
  author={Kaseb, Zeynab and M{\"o}ller, Matthias and Vergara, Pedro P and Palensky, Peter},
  journal={Scientific Reports},
  volume={14},
  number={1},
  note={{A}rt. no. 23216},
  year={2024},
  publisher={Nature Publishing Group UK London}
}

@article{ellinas2024hybrid,
  title={A hybrid quantum--classical algorithm for mixed-integer optimization in power systems},
  author={Ellinas, Petros and Chevalier, Samuel and Chatzivasileiadis, Spyros},
  journal={Electric Power Systems Research},
  volume={235},
  note={{A}rt. no. 110835},
  month=oct,
  year={2024},
  publisher={Elsevier}
}

@article{mahroo2023learning,
  title={Learning infused quantum-classical distributed optimization technique for power generation scheduling},
  author={Mahroo, Reza and Kargarian, Amin},
  journal={IEEE Transactions on Quantum Engineering},
  volume={4},
  pages={1--14},
  note={{A}rt. no. 3102314},
  year={2023},
  publisher={IEEE}
}

@article{morstyn2022annealing,
  title={Annealing-based quantum computing for combinatorial optimal power flow},
  author={Morstyn, Thomas},
  journal={IEEE Transactions on Smart Grid},
  volume={14},
  number={2},
  pages={1093--1102},
  month=mar,
  year={2023},
  publisher={IEEE}
}

@article{amani2025optimal,
  title={Optimal power flow solution via noise-resilient quantum interior-point methods},
  author={Amani, Farshad and Kargarian, Amin},
  journal={Electric Power Systems Research},
  volume={240},
  note={{A}rt. no. 111216},
  month=mar,
  year={2025},
  publisher={Elsevier}
}

@inproceedings{hafshejani2025quantum,
  title={Quantum algorithms for optimal power flow},
  author={Hafshejani, S Fathi and Uddin, Md Mohsin and Neufeld, David and Benkoczi, Robert and Gaur, Daya},
  booktitle={Proceedings of the 15th International Conference on Power, Energy, and Electrical Engineering (CPEEE)},
  pages={295--300},
  year={2025}
}

@article{fu2023coordinated,
  title={Coordinated post-disaster restoration for resilient urban distribution systems: A hybrid quantum-classical approach},
  author={Fu, Wei and Xie, Haipeng and Zhu, Hao and Wang, Hefeng and Jiang, Lizhou and Chen, Chen and Bie, Zhaohong},
  journal={Energy},
  volume={284},
  note={{A}rt. no. 129314},
  month=dec,
  year={2023},
  publisher={Elsevier}
}

@article{nikmehr2024quantum,
  title={Quantum annealing-infused microgrids formation: Distribution system restoration and resilience enhancement},
  author={Nikmehr, Nima and Zhang, Peng and Zheng, Honghao and Wei, Tzu-Chieh and He, Gang and Shamash, Yacov A},
  journal={IEEE Transactions on Power Systems},
  volume={40},
  number={1},
  pages={463--475},
  month=jan,
  year={2025},
  publisher={IEEE}
}

@article{bucher2024evaluating,
  title={Evaluating quantum optimization for dynamic self-reliant community detection},
  author={Bucher, David and Porawski, Daniel and Wimmer, Benedikt and N{\"u}{\ss}lein, Jonas and O’Meara, Corey and Mohseni, Naeimeh and Cortiana, Giorgio and Linnhoff-Popien, Claudia},
  journal={IEEE Transactions on Smart Grid},
  volume={16},
  number={2},
  pages={1339--1350},
  month=mar,
  year={2025},
  publisher={IEEE}
}

@article{wang2022quantum,
  title={Quantum annealing with integer slack variables for grid partitioning},
  author={Wang, Dawei and Zheng, Kedi and Teng, Fei and Chen, Qixin},
  journal={IEEE Transactions on Power Systems},
  volume={38},
  number={2},
  pages={1747--1750},
  month=mar,
  year={2023},
  publisher={IEEE}
}

@inproceedings{jing2022interoperation,
  title={Interoperation analysis of reconfigurable networked microgrids through quantum approximate optimization algorithm},
  author={Jing, Hang and Wang, Ye and Li, Yan},
  booktitle={Proceedings of the IEEE Power \& Energy Society General Meeting (PESGM)},
  pages={1--5},
  year={2022}
}

@article{jing2023data,
  title={Data-driven quantum approximate optimization algorithm for power systems},
  author={Jing, Hang and Wang, Ye and Li, Yan},
  journal={Communications Engineering},
  volume={2},
  number={1},
  note={{A}rt. no. 12},
  year={2023},
  publisher={Nature Publishing Group UK London}
}

@article{colucci2023power,
  title={Power network optimization: A quantum approach},
  author={Colucci, Giuseppe and van der Linde, Stan and Phillipson, Frank},
  journal={IEEE Access},
  volume={11},
  pages={98926--98938},
  year={2023},
  publisher={IEEE}
}

@article{silva2022qubo,
  title={A {QUBO} formulation for minimum loss network reconfiguration},
  author={Silva, Filipe FC and Carvalho, Pedro MS and Ferreira, Lu{\'\i}s AFM and Omar, Yasser},
  journal={IEEE Transactions on Power Systems},
  volume={38},
  number={5},
  pages={4559--4571},
  month=sep,
  year={2023},
  publisher={IEEE}
}

@article{silva2023quantum,
  title={A quantum computing approach for minimum loss problems in electrical distribution networks},
  author={Silva, Filipe FC and Carvalho, Pedro MS and Ferreira, Lu{\'\i}s AFM},
  journal={Scientific Reports},
  volume={13},
  number={1},
  note={{A}rt. no. 10777},
  year={2023},
  publisher={Nature Publishing Group UK London}
}

@article{bi2023quantum,
  title={Quantum annealing algorithm for fault section location in distribution networks},
  author={Bi, Zhongqin and Yang, Xiaoting and Wang, Baonan and Zhang, Weina and Dong, Zhen and Zhang, Dan},
  journal={Applied Soft Computing},
  volume={149},
  note={{A}rt. no. 110973},
  month=dec,
  year={2023},
  publisher={Elsevier}
}

@article{fei2024power,
  title={Power system fault diagnosis with quantum computing and efficient gate decomposition},
  author={Fei, Xiang and Zhao, Huan and Zhou, Xiyuan and Zhao, Junhua and Shu, Ting and Wen, Fushuan},
  journal={Scientific Reports},
  volume={14},
  number={1},
  note={{A}rt. no. 16991},
  year={2024},
  publisher={Nature Publishing Group UK London}
}

@article{kea2023leveraging,
  title={Leveraging knapsack {QAOA} approach for optimal electric vehicle charging},
  author={Kea, Kimleang and Huot, Chansreynich and Han, Youngsun},
  journal={IEEE Access},
  volume={11},
  pages={109964--109973},
  year={2023},
  publisher={IEEE}
}

@article{li2025quantum,
  title={Quantum assisted combinatorial {B}enders' algorithm for the synergy of hydrogen and power distribution systems with mobile storage},
  author={Li, Mingze and Wang, Siyuan and Fan, Lei and Han, Zhu},
  journal={IEEE Transactions on Power Systems},
  volume={40},
  number={5},
  pages={3967--3978},
  month=sep,
  year={2025},
  publisher={IEEE}
}

@article{ajagekar2024variational,
  title={Variational quantum circuit based demand response in buildings leveraging a hybrid quantum-classical strategy},
  author={Ajagekar, Akshay and You, Fengqi},
  journal={Applied Energy},
  volume={364},
  note={{A}rt. no. 123244},
  month=jun,
  year={2024},
  publisher={Elsevier}
}

@article{bucher2025grid,
  title={Grid Cost Allocation in Peer-to-Peer Electricity Markets: Benchmarking Classical and Quantum Optimization Approaches},
  author={Bucher, David and Porawski, Daniel and Wimmer, Benedikt and N{\"u}{\ss}lein, Jonas and O'Meara, Corey and Cortiana, Giorgio and Linnhoff-Popien, Claudia},
  journal={arXiv preprint arXiv:2501.05253},
  year={2025}
}

@inproceedings{han2024quantum,
  title={Quantum-assisted stochastic economic dispatch for renewables-rich power systems},
  author={Han, Xutao and Li, Zhiyi and Xu, Yue},
  booktitle={Proceedings of the IEEE Power \& Energy Society General Meeting (PESGM)},
  pages={1--5},
  year={2024}
}

@book{saad2003iterative,
  title={Iterative Methods for Sparse Linear Systems},
  author={Saad, Yousef},
  year={2003},
  publisher={SIAM}
}

@article{wang2015emtp,
  title={{EMTP}-type program realization of {K}rylov subspace based model reduction methods for large-scale active distribution network},
  author={Wang, Chengshan and Yu, Hao and Li, Peng and Wu, Jianzhong and Ding, Chengdi and Song, Guanyu and Lin, Dun and Xing, Feng},
  journal={CSEE Journal of Power and Energy Systems},
  volume={1},
  number={1},
  pages={52--60},
  month=mar,
  year={2015},
  publisher={CSEE}
}

@article{hosaka2023preconditioning,
  title={Preconditioning for a variational quantum linear solver},
  author={Hosaka, Aruto and Yanagisawa, Koichi and Koshikawa, Shota and Kudo, Isamu and Alifu, Xiafukaiti and Yoshida, Tsuyoshi},
  journal={arXiv preprint arXiv:2312.15657},
  year={2023}
}

@article{turati2024empirical,
  title={An empirical analysis on the effectiveness of the variational quantum linear solver},
  author={Turati, Gloria and Marruzzo, Alessia and Dacrema, Maurizio Ferrari and Cremonesi, Paolo},
  journal={arXiv preprint arXiv:2409.06339},
  year={2024}
}

@article{cai2023quantum,
  title={Quantum error mitigation},
  author={Cai, Zhenyu and Babbush, Ryan and Benjamin, Simon C and Endo, Suguru and Huggins, William J and Li, Ying and McClean, Jarrod R and O’Brien, Thomas E},
  journal={Reviews of Modern Physics},
  volume={95},
  number={4},
  note={{A}rt. no. 045005},
  month=dec,
  year={2023},
  publisher={APS}
}

@article{devitt2013quantum,
  title={Quantum error correction for beginners},
  author={Devitt, Simon J and Munro, William J and Nemoto, Kae},
  journal={Reports on Progress in Physics},
  volume={76},
  number={7},
  note={{A}rt. no. 076001},
  year={2013},
  publisher={IOP Publishing}
}

@article{chen2024enabling,
  title={Enabling large-scale and high-precision fluid simulations on near-term quantum computers},
  author={Chen, Zhao-Yun and Ma, Teng-Yang and Ye, Chuang-Chao and Xu, Liang and Bai, Wen and Zhou, Lei and Tan, Ming-Yang and Zhuang, Xi-Ning and Xu, Xiao-Fan and Wang, Yun-Jie and others},
  journal={Computer Methods in Applied Mechanics and Engineering},
  volume={432},
  note={{A}rt. no. 117428},
  month=dec,
  year={2024},
  publisher={Elsevier}
}

@article{saghafi2025multifractality,
  title={Multifractality Analysis of Single Qubit Quantum Circuit Outcomes for a Superconducting Quantum Computer},
  author={Saghafi, Mohammadreza and Mili, Lamine and Wirsing, Karlton},
  journal={arXiv preprint arXiv:2512.18491},
  year={2025}
}

@article{abdi2010principal,
  title={Principal Component Analysis},
  author={Abdi, Herv{\'e} and Williams, Lynne J},
  journal={Wiley Interdisciplinary Reviews: Computational Statistics},
  volume={2},
  number={4},
  pages={433--459},
  year={2010},
  publisher={Wiley Online Library}
}

@incollection{suthaharan2016support,
  title={Support Vector Machine},
  author={Suthaharan, Shan},
  booktitle={Machine Learning Models and Algorithms for Big Data Classification: Thinking with Examples for Effective Learning},
  pages={207--235},
  year={2016},
  publisher={Springer}
}

@article{lloyd2014quantum,
  title={Quantum principal component analysis},
  author={Lloyd, Seth and Mohseni, Masoud and Rebentrost, Patrick},
  journal={Nature Physics},
  volume={10},
  number={9},
  pages={631--633},
  year={2014},
  publisher={Nature Publishing Group UK London}
}

@article{xin2021experimental,
  title={Experimental quantum principal component analysis via parametrized quantum circuits},
  author={Xin, Tao and Che, Liangyu and Xi, Cheng and Singh, Amandeep and Nie, Xinfang and Li, Jun and Dong, Ying and Lu, Dawei},
  journal={Physical Review Letters},
  volume={126},
  number={11},
  note={{A}rt. no. 110502},
  month=mar,
  year={2021},
  publisher={APS}
}

@article{rebentrost2014quantum,
  title={Quantum support vector machine for big data classification},
  author={Rebentrost, Patrick and Mohseni, Masoud and Lloyd, Seth},
  journal={Physical Review Letters},
  volume={113},
  number={13},
  note={{A}rt. no. 130503},
  month=sep,
  year={2014},
  publisher={APS}
}

@inproceedings{ding2024zc,
  title={Zc (3900) observation at BESIII with QSVM method},
  author={Ding, Biao and Meng, Zhaoxia and Zou, Jiaheng and Li, Teng and Lin, Tao},
  booktitle={EPJ Web of Conferences},
  volume={295},
  note={{A}rt. no. 12008},
  year={2024},
  organization={EDP Sciences}
}

@article{an2020data,
  title={A data-driven method for transient stability margin prediction based on security region},
  author={An, Jun and Yu, Jiachen and Li, Zonghan and Zhou, Yibo and Mu, Gang},
  journal={Journal of Modern Power Systems and Clean Energy},
  volume={8},
  number={6},
  pages={1060--1069},
  month=nov,
  year={2020},
  publisher={SGEPRI}
}

@article{tian2021neural,
  title={Neural-network-based power system state estimation with extended observability},
  author={Tian, Guanyu and Gu, Yingzhong and Shi, Di and Fu, Jing and Yu, Zhe and Zhou, Qun},
  journal={Journal of Modern Power Systems and Clean Energy},
  volume={9},
  number={5},
  pages={1043--1053},
  month=sep,
  year={2021},
  publisher={SGEPRI}
}

@article{yu2023provable,
  title={Provable advantage of parameterized quantum circuit in function approximation},
  author={Yu, Zhan and Chen, Qiuhao and Jiao, Yuling and Li, Yinan and Lu, Xiliang and Wang, Xin and Yang, Jerry Zhijian},
  journal={arXiv preprint arXiv:2310.07528},
  year={2023}
}

@article{zhou2022noise,
  title={Noise-resilient quantum machine learning for stability assessment of power systems},
  author={Zhou, Yifan and Zhang, Peng},
  journal={IEEE Transactions on Power Systems},
  volume={38},
  number={1},
  pages={475--487},
  month=jan,
  year={2023},
  publisher={IEEE}
}

@article{sebastianelli2021circuit,
  title={On circuit-based hybrid quantum neural networks for remote sensing imagery classification},
  author={Sebastianelli, Alessandro and Zaidenberg, Daniela Alessandra and Spiller, Dario and Le Saux, Bertrand and Ullo, Silvia Liberata},
  journal={IEEE Journal of Selected Topics in Applied Earth Observations and Remote Sensing},
  volume={15},
  pages={565--580},
  year={2021},
  publisher={IEEE}
}

@article{wang2024shallow,
  title={Shallow hybrid quantum-classical convolutional neural network model for image classification},
  author={Wang, Aijuan and Hu, Jianglong and Zhang, Shiyue and Li, Lusi},
  journal={Quantum Information Processing},
  volume={23},
  number={1},
  pages={17},
  year={2024},
  publisher={Springer}
}

@article{gong2024quantum,
  title={Quantum convolutional neural network based on variational quantum circuits},
  author={Gong, Li-Hua and Pei, Jun-Jie and Zhang, Tian-Feng and Zhou, Nan-Run},
  journal={Optics Communications},
  volume={550},
  note={{A}rt. no. 129993},
  month=jan,
  year={2024},
  publisher={Elsevier}
}

@article{dallaire2018quantum,
  title={Quantum generative adversarial networks},
  author={Dallaire-Demers, Pierre-Luc and Killoran, Nathan},
  journal={arXiv preprint arXiv:1804.08641},
  year={2018}
}

@article{lloyd2018quantum,
  title={Quantum generative adversarial learning},
  author={Lloyd, Seth and Weedbrook, Christian},
  journal={Physical Review Letters},
  volume={121},
  number={4},
  note={{A}rt. no. 040502},
  month=jul,
  year={2018},
  publisher={APS}
}

@article{situ2020quantum,
  title={Quantum generative adversarial network for generating discrete distribution},
  author={Situ, Haozhen and He, Zhimin and Wang, Yuyi and Li, Lvzhou and Zheng, Shenggen},
  journal={Information Sciences},
  volume={538},
  pages={193--208},
  month=oct,
  year={2020},
  publisher={Elsevier}
}

@article{zoufal2019quantum,
  title={Quantum generative adversarial networks for learning and loading random distributions},
  author={Zoufal, Christa and Lucchi, Aur{\'e}lien and Woerner, Stefan},
  journal={npj Quantum Information},
  volume={5},
  number={1},
  pages={103},
  year={2019},
  publisher={Nature Publishing Group UK London}
}

@article{chakrabarti2019quantum,
  title={Quantum {W}asserstein generative adversarial networks},
  author={Chakrabarti, Shouvanik and Yiming, Huang and Li, Tongyang and Feizi, Soheil and Wu, Xiaodi},
  journal={Advances in Neural Information Processing Systems},
  volume={32},
  year={2019}
}

@article{zhu2022generative,
  title={Generative quantum learning of joint probability distribution functions},
  author={Zhu, Elton Yechao and Johri, Sonika and Bacon, Dave and Esencan, Mert and Kim, Jungsang and Muir, Mark and Murgai, Nikhil and Nguyen, Jason and Pisenti, Neal and Schouela, Adam and others},
  journal={Physical Review Research},
  volume={4},
  number={4},
  note={{A}rt. no. 043092},
  year={2022},
  publisher={APS}
}

@article{tsang2023hybrid,
  title={Hybrid quantum--classical generative adversarial network for high-resolution image generation},
  author={Tsang, Shu Lok and West, Maxwell T and Erfani, Sarah M and Usman, Muhammad},
  journal={IEEE Transactions on Quantum Engineering},
  volume={4},
  pages={1--19},
  year={2023},
  note={{A}rt no. 3102419},
  publisher={IEEE}
}

@article{horowitz2022quantum,
  title={A quantum generative model for multi-dimensional time series using Hamiltonian learning},
  author={Horowitz, Haim and Rao, Pooja and Radha, Santosh Kumar},
  journal={arXiv preprint arXiv:2204.06150},
  year={2022}
}

@article{tucci1997quantum,
  title={Quantum {B}ayesian nets},
  author={Tucci, Robert R},
  journal={arXiv preprint quant-ph/9706039},
  year={1997}
}

@article{low2014quantum,
  title={Quantum inference on {B}ayesian networks},
  author={Low, Guang Hao and Yoder, Theodore J and Chuang, Isaac L},
  journal={Physical Review A},
  volume={89},
  number={6},
  note={{A}rt no. 062315},
  year={2014},
  publisher={APS}
}

@article{berner2021quantum,
  title={Quantum {B}ayesian neural networks},
  author={Berner, Noah and Fortuin, Vincent and Landman, Jonas},
  journal={arXiv preprint arXiv:2107.09599},
  year={2021}
}

@article{amin2018quantum,
  title={Quantum {B}oltzmann machine},
  author={Amin, Mohammad H and Andriyash, Evgeny and Rolfe, Jason and Kulchytskyy, Bohdan and Melko, Roger},
  journal={Physical Review X},
  volume={8},
  number={2},
  note={{A}rt. no. 021050},
  year={2018},
  publisher={APS}
}

@article{zoufal2021variational,
  title={Variational quantum {B}oltzmann machines},
  author={Zoufal, Christa and Lucchi, Aur{\'e}lien and Woerner, Stefan},
  journal={Quantum Machine Intelligence},
  volume={3},
  number={1},
  pages={7},
  year={2021},
  publisher={Springer}
}

@article{coopmans2024sample,
  title={On the sample complexity of quantum {B}oltzmann machine learning},
  author={Coopmans, Luuk and Benedetti, Marcello},
  journal={Communications Physics},
  volume={7},
  number={1},
  note={{A}rt. no. 274},
  year={2024},
  publisher={Nature Publishing Group UK London}
}

@article{torlai2017neural,
  title={Neural decoder for topological codes},
  author={Torlai, Giacomo and Melko, Roger G},
  journal={Physical Review Letters},
  volume={119},
  number={3},
  note={{A}rt. no. 030501},
  month=jul,
  year={2017},
  publisher={APS}
}

@article{perdomo2018opportunities,
  title={Opportunities and challenges for quantum-assisted machine learning in near-term quantum computers},
  author={Perdomo-Ortiz, Alejandro and Benedetti, Marcello and Realpe-G{\'o}mez, John and Biswas, Rupak},
  journal={Quantum Science and Technology},
  volume={3},
  number={3},
  note={{A}rt. no. 030502},
  year={2018},
  publisher={IOP Publishing}
}

@inproceedings{zhang2025llm,
  title={{LLM-QUBO}: An End-to-End Framework for Automated {QUBO} Transformation from Natural Language Problem Descriptions},
  author={Zhang, Huixiang and Emu, Mahzabeen and Choudhury, Salimur},
  booktitle={Proceedings of the AAAI Symposium Series},
  volume={7},
  number={1},
  pages={411--418},
  year={2025}
}

@article{liu2025quantum,
  title={Quantum relational knowledge distillation},
  author={Liu, Chen-Yu and Chen, Kuan-Cheng and Murota, Keisuke and Chen, Samuel Yen-Chi and Rinaldi, Enrico},
  journal={arXiv preprint arXiv:2508.13054},
  year={2025}
}

@article{de2021materials,
  title={Materials challenges and opportunities for quantum computing hardware},
  author={De Leon, Nathalie P and Itoh, Kohei M and Kim, Dohun and Mehta, Karan K and Northup, Tracy E and Paik, Hanhee and Palmer, BS and Samarth, Nitin and Sangtawesin, Sorawis and Steuerman, David W},
  journal={Science},
  volume={372},
  number={6539},
  note={{A}rt. no. eabb2823},
  year={2021},
  publisher={American Association for the Advancement of Science}
}

@BOOK{NAP25196,
  author    = {{National Academies of Sciences, Engineering, and Medicine}},
  editor    = {Emily Grumbling and Mark Horowitz},
  title     = {Quantum Computing: Progress and Prospects},
  isbn      = {978-0-309-47969-1},
  doi       = {10.17226/25196},
  year      = {2019},
  publisher = {The National Academies Press}
}

@misc{intel2023chip,
  author       = {{Intel Corporation}},
  title        = {Intel's New Chip to Advance Silicon Spin Qubit Research for Quantum Computing},
  howpublished = {Intel Newsroom},
  year         = {2023},
  month        = jun,
  day          = 15,
  url          = {https://newsroom.intel.com/new-technologies/quantum-computing-chip.}
}

@misc{microsoft2024topological,
  author       = {{Microsoft Corporation}},
  title        = {Topological qubits},
  howpublished = {Microsoft Quantum},
  year         = {2024}, 
  url          = {
https://quantum.microsoft.com/en-us/insights/education/concepts/topolo
 gical-qubits.}
}

@article{golec2024quantum,
  title={Quantum cloud computing: Trends and challenges},
  author={Golec, Muhammed and Hatay, Emir Sahin and Golec, Mustafa and Uyar, Murat and Golec, Merve and Gill, Sukhpal Singh},
  journal={Journal of Economy and Technology},
  volume={2},
  pages={190--199},
  year={2024},
  publisher={Elsevier}
}

@misc{ibmquantum,
  author       = {{IBM Corporation}},
  title        = {IBM Quantum Platform},
  howpublished = {IBM Quantum},
  url          = {https://quantum.cloud.ibm.com/.}
}

@misc{google2025quantum,
  author       = {{Google Quantum AI}},
  title        = {Quantum {AI}},
  howpublished = {Google},
  year         = {2025},
  url          = {https://quantumai.google/.}
}

@misc{rigetti_qcs,
  author       = {{Rigetti Computing Inc.}},
  title        = {Quantum Cloud Services ({QCS})},
  howpublished = {Rigetti QCS},
  url          = {https://qcs.rigetti.com/.}
}

@misc{aws_braket,
  author       = {{Amazon Web Services, Inc.}},
  howpublished = {AWS Quantum Technologies},
  url          = {https://aws.amazon.com/braket/.}
}

@misc{microsoft_azure_quantum,
  author       = {{Microsoft Corporation}},
  title        = {Azure Quantum},
  howpublished = {Microsoft Azure},
  url          = {https://azure.microsoft.com/en-us/products/quantum/.}
}

@misc{dwave_leap,
  author       = {{D-Wave Systems Inc.}},
  title        = {Leap Quantum Cloud Service},
  howpublished = {D-Wave},
  url          = {https://cloud.dwavesys.com/leap/.}
}

@misc{xanadu_cloud,
  author       = {{Xanadu Quantum Technologies Inc.}},
  title        = {Xanadu Cloud},
  howpublished = {Xanadu},
  url          = {https://cloud.xanadu.ai/.}
}

@misc{qutech_quantum_inspire,
  author       = {{QuTech}},
  title        = {Quantum Inspire},
  howpublished = {QuTech Quantum AI Lab},
  url          = {https://www.quantum-inspire.com/.}
}

@misc{oqc_qcaas,
  author       = {{Oxford Quantum Circuits Ltd.}},
  title        = {Application Optimised Quantum Compute},
  howpublished = {Oxford Quantum Circuits (OQC)},
  year         = {2021},
  url          = {https://oqc.tech/.}
}

@misc{pasqal_cloud,
  author       = {{Pasqal}},
  title        = {Welcome to {Pasqal} Cloud Services},
  howpublished = {Pasqal Cloud Services Documentation},
  url          = {https://docs.pasqal.com/cloud/.}
}

@misc{quandela_cloud,
  author       = {{Quandela}},
  howpublished = {Quandela Cloud},
  url          = {https://cloud.quandela.com/.}
}

@misc{ctq_tianyan,
  author       = {{China Telecom Quantum Technology Co., Ltd.}},
  title        = {Tianyan Quantum Cloud Platform},
  howpublished = {China Telecom},
  url          = {https://qc.zdxlz.com/quantumAdvantage?lang=en.}
}

@misc{quantumctek_cloud_en,
  author       = {{QuantumCTek Co., Ltd.}},
  title        = {QuantumCTek Quantum Computing Cloud Platform},
  howpublished = {QuantumCTek},
  url          = {https://quantumctek-cloud.com/.}
}

@misc{baqis_quafu,
  author       = {{Beijing Academy of Quantum Information Sciences}},
  title        = {Quafu Cloud Platform},
  howpublished = {BAQIS},
  url          = {https://quafu.baqis.ac.cn/#/home.},
}

@misc{arclight_cloud,
  author       = {{Arclight Quantum Computing Inc.}},
  howpublished = {Arclight Quantum},
  url          = {https://qcloud.arclightquantum.com/.}
}

@misc{origin_cloud,
  author       = {{Origin Quantum Co., Ltd.}},
  title        = {Origin Quantum Cloud Platform},
  howpublished = {Origin Quantum},
  url          = {https://qcloud.originqc.com.cn/en/home.}
}

@misc{huayi_cloud,
  author       = {{Huayi Quantum Technology Co., Ltd.}},
  howpublished = {Huayi Quantum},
  url          = {https://www.hyqubit.com/simulator.},
}

@misc{spinq_cloud,
  author       = {{Shenzhen SpinQ Technology Co., Ltd.}},
  howpublished = {SpinQ Cloud},
  url          = {https://cloud.spinq.cn/.},
}

@misc{cmcc_wuyue_quantum,
  author       = {{China Mobile}},
  url          = {https://ecloud.10086.cn/portal/product/WYQCLOUD.}
}

@misc{qboson_cloud,
  author       = {{Beijing QBoson Quantum Technology Co., Ltd.}},
  howpublished = {QBoson},
  url          = {https://platform.qboson.com/.}
}

@article{chen2023complexity,
  title={The complexity of {NISQ}},
  author={Chen, Sitan and Cotler, Jordan and Huang, Hsin-Yuan and Li, Jerry},
  journal={Nature Communications},
  volume={14},
  number={1},
  note={{A}rt. no. 6001},
  year={2023},
  publisher={Nature Publishing Group UK London}
}

@article{larocca2025barren,
  title={Barren plateaus in variational quantum computing},
  author={Larocca, Martin and Thanasilp, Supanut and Wang, Samson and Sharma, Kunal and Biamonte, Jacob and Coles, Patrick J and Cincio, Lukasz and McClean, Jarrod R and Holmes, Zo{\"e} and Cerezo, Marco},
  journal={Nature Reviews Physics},
  volume={7},
  number={4},
  pages={174--189},
  year={2025},
  publisher={Nature Publishing Group UK London}
}

@book{glover2002power,
  title={Power System Analysis and Design},
  author={Glover, J Duncan and Overbye, Thomas J and Sarma, Mulukutla S and Birchfield, Adam B},
  year={2023},
  publisher={Cengage Learning},
  edition = {7}
}

@article{siegl2024solving,
  title={Solving Transport Equations on Quantum Computers-Potential and Limitations of Physics-Informed Quantum Circuits},
  author={Siegl, Pia and Wassing, Simon and Mieth, Dirk Markus and Langer, Stefan and Bekemeyer, Philipp},
  journal={CEAS Aeronautical Journal},
  year={2024},
  publisher={Springer}
}

@article{du2020expressive,
  title={Expressive power of parametrized quantum circuits},
  author={Du, Yuxuan and Hsieh, Min-Hsiu and Liu, Tongliang and Tao, Dacheng},
  journal={Physical Review Research},
  volume={2},
  number={3},
  note={{A}rt. no. 033125},
  year={2020},
  publisher={APS}
}

@article{abbas2021power,
  title={The power of quantum neural networks},
  author={Abbas, Amira and Sutter, David and Zoufal, Christa and Lucchi, Aur{\'e}lien and Figalli, Alessio and Woerner, Stefan},
  journal={Nature Computational Science},
  volume={1},
  number={6},
  pages={403--409},
  year={2021},
  publisher={Nature Publishing Group US New York}
}

@article{huang2025generative,
  title={Generative quantum advantage for classical and quantum problems},
  author={Huang, Hsin-Yuan and Broughton, Michael and Eassa, Norhan and Neven, Hartmut and Babbush, Ryan and McClean, Jarrod R},
  journal={arXiv preprint arXiv:2509.09033},
  year={2025}
}

@ARTICLE{11342337,
  author={Yao, Yongbing and Xu, Yijun and Gu, Wei and Wang, Yishen and Liu, Chengjun and Lu, Shuai and Mili, Lamine},
  journal={IEEE Transactions on Sustainable Energy}, 
  title={Renewable Scenario Generation Using Quantum Generative Adversarial Networks}, 
  year={2026},
  volume={},
  number={},
  pages={1-15},
  note={early access},
  doi={10.1109/TSTE.2026.3652012}}

@article{wang2025hybrid,
  title={Hybrid Quantum-Classical Inception Neural Network for Image Classification},
  author={Wang, Lu and Liu, Yuxiang and Ji, Jinpeng and Meng, Fanxu and Zhang, Zaichen and Yu, Xutao},
  journal={Advanced Quantum Technologies},
  volume={8},
  number={10},
  note={{A}rt. no. 2400700},
  year={2025},
  publisher={Wiley Online Library}
}

@misc{qiskit2024,
      title={Quantum computing with {Q}iskit},
      author={Javadi-Abhari, Ali and Treinish, Matthew and Krsulich, Kevin and Wood, Christopher J. and Lishman, Jake and Gacon, Julien and Martiel, Simon and Nation, Paul D. and Bishop, Lev S. and Cross, Andrew W. and Johnson, Blake R. and Gambetta, Jay M.},
      year={2024},
      doi={10.48550/arXiv.2405.08810},
      eprint={2405.08810},
      archivePrefix={arXiv},
      primaryClass={quant-ph}
}

@article{zhang2023tensorcircuit,
  title={Tensorcircuit: A quantum software framework for the {NISQ} era},
  author={Zhang, Shi-Xin and Allcock, Jonathan and Wan, Zhou-Quan and Liu, Shuo and Sun, Jiace and Yu, Hao and Yang, Xing-Han and Qiu, Jiezhong and Ye, Zhaofeng and Chen, Yu-Qin and others},
  journal={Quantum},
  volume={7},
  note={{A}rt. no. 912},
  year={2023},
  publisher={Verein zur F{\"o}rderung des Open Access Publizierens in den Quantenwissenschaften}
}

@article{bergholm2018pennylane,
  title={Pennylane: Automatic differentiation of hybrid quantum-classical computations},
  author={Bergholm, Ville and Izaac, Josh and Schuld, Maria and Gogolin, Christian and Ahmed, Shahnawaz and Ajith, Vishnu and Alam, M Sohaib and Alonso-Linaje, Guillermo and AkashNarayanan, Bharath and Asadi, Ali and others},
  journal={arXiv preprint arXiv:1811.04968},
  year={2018}
}

@article{blei2017variational,
  title={Variational inference: A review for statisticians},
  author={Blei, David M and Kucukelbir, Alp and McAuliffe, Jon D},
  journal={Journal of the American statistical Association},
  volume={112},
  number={518},
  pages={859--877},
  year={2017},
  publisher={Taylor \& Francis}
}

@article{saghafi2025predictive,
  title={Predictive time-series analysis of single-qubit quantum circuit outcomes for a superconducting quantum computer: Forecasting error patterns},
  author={Saghafi, Mohammadreza and Mili, Lamine},
  journal={IEEE Access},
  volume={13},
  pages={40115--40132},
  year={2025},
  publisher={IEEE}
}

\end{document}